\documentclass{article}

\usepackage[style=numeric, sorting=none, doi=true, maxcitenames=3, mincitenames=1]{biblatex}
\usepackage{authblk}
\usepackage[a4paper, total={7in, 10in}]{geometry}
\usepackage{amsmath, amssymb}
\usepackage{graphicx}
\usepackage{float}
\usepackage{hyperref}
\usepackage{subcaption}
\usepackage{multirow}
\usepackage{algorithm}
\usepackage{tikz}
\usepackage[noend]{algpseudocode}

\DeclareMathOperator*{\argmax}{arg\,max}

\counterwithin{figure}{section}
\counterwithin{table}{section}
\counterwithin{algorithm}{section}

\renewenvironment{abstract}{%
	\begin{center}
		\begin{minipage}{0.75\textwidth}
			\footnotesize
			\textbf{Abstract: }
		}{%
		\end{minipage}
	\end{center}
	\vspace{5mm}
}

\title{Optimising Geometric Deep Learning for Varied Reconstruction Tasks in High Granularity Calorimeters}

\author[1]{Matthieu Melennec}
\author[1]{Fr\'ed\'eric Magniette}

\affil[1]{Ecole Polytechnique, IN2P3-CNRS, Laboratoire Leprince-Ringuet, F-91120 Palaiseau, France}

\date{}

\begin{document}
	
	\maketitle
	
	\begin{abstract}
		In the recent years, high energy physics discoveries have been driven by the increasing of luminosity and/or detector granularity. This evolution gives access to bigger statistics and data samples, but can make it hard to process the detector outputs with current methods and algorithms. Graph Neural Networks (GNNs), have been shown to be powerful tools to address these challenges. While GNNs can deal with the non-euclidean nature of the data extracted from these detectors, their deployment presents significant difficulties, mainly from the computational complexity of the algorithms they are based on.
		
		We propose to use the known geometries of the detectors to optimise the algorithms used for GNN pipelines, which usually have quadratic algorithmic complexities. We validate the validity of our optimised approach on various tasks, such as particle identification, energy regression and instance segmentation. Finally, we present HIBOU, a generic modular architecture that provides a blueprint for how these GNN blocks are to be implemented within experiment-specific reconstruction frameworks.
	\end{abstract}
	
	
	\section{Introduction}
	
		The High-Luminosity LHC (HL-LHC) upgrade \cite{aberle_20} will induce severe pileup with nearly 200 simultaneous collisions per bunch crossing, causing spatial overlapping of energy deposits in conventional calorimeters. To overcome this limitation and disentangle individual particle showers, next-generation experiments are deploying highly granular detectors, such as the CMS High Granularity Calorimeter \cite{pasztor_23, cms_17, atlas_17}.
		
		To fully exploit the unprecedented level of precision offered by this high granularity image of the shower development, new algorithms must be developed \cite{cms_17}. While deep learning based reconstruction algorithms have shown their success on the conventional calorimeter geometries \cite{simkina_24, maidannyk_26}, they can be hard to generalise to high dimensional outputs with variable granularity, such as the future HGCAL.  Graph Neural Networks (GNN) \cite{gilmer_17} are promising candidates for particle reconstruction in these complex environments \cite{qasim_19, bhattacharya_23}. However, they rely on algorithms that are prohibitively expensive to use. For instance, graph construction can usually only be done in a quadratic number of operations in the number of nodes \cite{qasim_19, franceschi_19}, and is often used repeatedly within the same pipeline.
		
		In this context, the OGCID project (Optimised Graph Convolution for particle IDentification) aims at optimising the algorithms that are the building blocks of GNN pipelines, to work towards their generalised deployment. The objectives of the OGCID project include the optimisation of graph construction algorithms, the optimisation of graph convolution network operations applied to detector data and the parallelisation of these algorithms. OGCID aims to validate these algorithms with various tasks in a simple detector, used as an algorithmic testbed, to provide an architectural blueprint for the instantiation of GNNs within experiment-specific deconstruction frameworks.
		
		In this paper, we present an algorithmic framework of Graph Neural Networks, most particularly when applied to particle detectors. We present implementations of optimised GNN pipeline operations, namely an optimised graph construction framework, by pre-computing the proximities of sensors, as well as optimised graph pooling operations that respect the structural information of the signals. We also present how this framework can be parallelised. 
		
		We confront this framework with particle reconstruction tasks in a high-granular sampling calorimeter. In particular, we design OPIF, a pipeline for particle identification from raw hits in the detector. We also design a model for the regression of the energy of particles, the ERGO model. It can be specialised to different particle types, based on the known typical signatures of particle decays in matter. A parallelised implementation of ERGO is studied. We also present a GNN based segmentation framework, the SOOP pipeline. It reconstructs individual showers from a pair of overlapping objects in the detector, which may then be exploited individually.
		
		These task-specific pipelines are finally synthesised into HIBOU, a generic modular architecture showing how the validated components can be organised and integrated into detector-specific reconstruction frameworks.
	
	\section{Detector and Event Simulation}
		
		The data used throughout this work is extracted from the D2 dataset \cite{becheva_24}. It is a set of particle interaction with a simplified high granularity calorimeter	inspired by CMS HGCAL \cite{cms_17_hgcal_tdr}, simulated with \textsc{g\'eant}4.
		
		D2 is used as a controlled testbed for the development and evaluation of detector-independent algorithmic principles. It is not intended to reproduce the complete reconstruction environment or computing constraints of a specific experiment. On the contrary, D2 is used to establish an upper bound for the attainable performance of our model in a calorimeter.
		
		\subsection{The Detector Structure}
		
\begin{figure}
				\centering
				\begin{subfigure}{0.45\textwidth}
					\centering
					\includegraphics[width=\textwidth]{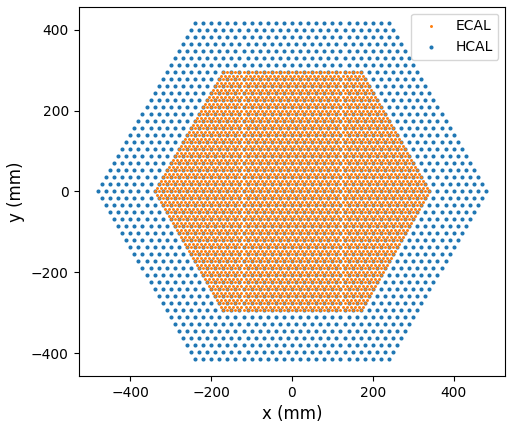}
				\end{subfigure}
				\begin{subfigure}{0.51\textwidth}
					\centering
					\includegraphics[width=\textwidth]{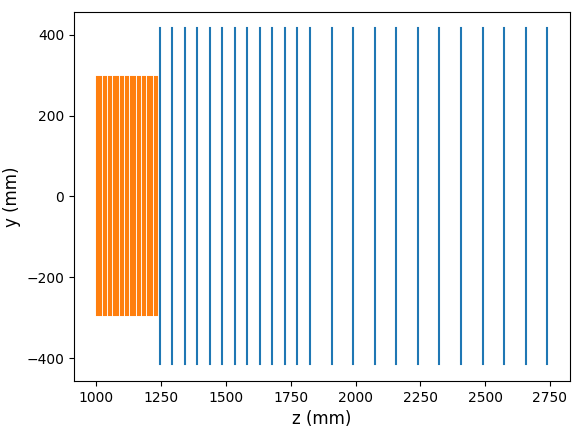}
				\end{subfigure}
				\caption{The D2 detector geometry. We observe the two sections of the detector, with the more granular ECAL (orange), followed by the HCAL (blue). We observe in particular that the HCAL has two subsections with different longitudinal granularity.}\label{sec:d2-fig:geom}
			\end{figure}

			The D2 detector is divided in two sections: an electromagnetic calorimeter (ECAL) of 26 layers, followed by a hadronic calorimeter (HCAL), itself divided into two subsections of 12 layers each: the forward HCAL, with a higher longitudinal granularity than the backward HCAL, further down from the interaction point. The full structure is represented in Figure \ref{sec:d2-fig:geom}.
			
			The detector consists of alternating pairs of absorber and silicon layers. The ECAL absorber layers are made of lead and have a thickness of 6.05 mm, while the HCAL section absorber layers made of stainless steel, the first 12 with thickness of 45mm and the second 12 with a thickness 80 mm. All active silicon layers are transversely segmented into hexagonal cells of 320 $\mu$m of thickness with surface of about 1 cm$^2$ for the ECAL and 4 cm$^2$ for the HCAL. The total transverse section of the ECAL has a diameter of 69 cm, while the HCAL has a diameter of 98 cm. These dimensions were chosen to ensure at least 95\% containment of the radial extension of the electromagnetic and hadronic showers. The pairs of absorber and active layers are placed at a distance of 3 mm each from other, and the total length of the calorimeter is 1.84 m with 0.24 m ($=28 X_0$) for the ECAL and 1.60 m ($=9\lambda$) for the HCAL. Table \ref{sec:d2-tab:structure} gives a summary of the detector's structure.
			\begin{table}
				\centering
				\begin{tabular}{l || l | l | l | l}
					\textbf{Parameter} & \textbf{ECAL} & \textbf{Near HCAL} & \textbf{Far HCAL} & \textbf{TOTAL} \\
					\hline\hline
					Nb layers & 26 & 12 & 12 & 50 \\
					Layer spacing & 3.00 mm & 3.00 mm & 3.00 mm & -- \\
					Cell material & Si & Si & Si & -- \\
					Cell thickness & 0.32 mm & 0.32 mm & 0.32 mm & -- \\
					Cell diameter & 10.0 mm & 20.0 mm & 20.0 mm & -- \\
					Absorber plate material & Pb & Stainless steel & Stainless steel & -- \\
					Absorber plate thickness & 6.05 mm & 45.0 mm & 80.0 mm & -- \\
					Length & 240.62 mm & 1576.68 mm & 820.46 mm & 1820.3 mm \\
					Radius & 345.0 mm & 490.0 mm & 490.0 mm & 490.0 mm
				\end{tabular}
				\caption{Calorimeter Specifications for the D2 detector.} \label{sec:d2-tab:structure}
			\end{table}
		
		\subsection{Particle Events}\label{chap:sim-sec:d2-subsec:events}
		
			Events are simulated using the \textsc{g\'eant}4 software \cite{allison_16}. The detector is centred around the $Oz$ axis, in a chamber containing air, with no external magnetic field. Events are sampled for electrons, photons, muons and positive pions, which are each representative of their class of particles. They shot by a particle gun one meter away from the first layer of the ECAL, on the $Oz$ axis. Energies range from 10 to 100 GeV for each particle type, and they are shot at an angle from the longitudinal axis ranging between $0^\circ$ and $15^\circ$. The azimuthal angle ranges uniformly between $0^\circ$ and $360^\circ$.
			
\begin{figure}
				\centering
				\begin{subfigure}{0.48\textwidth}
					\centering
					\includegraphics[width=\linewidth]{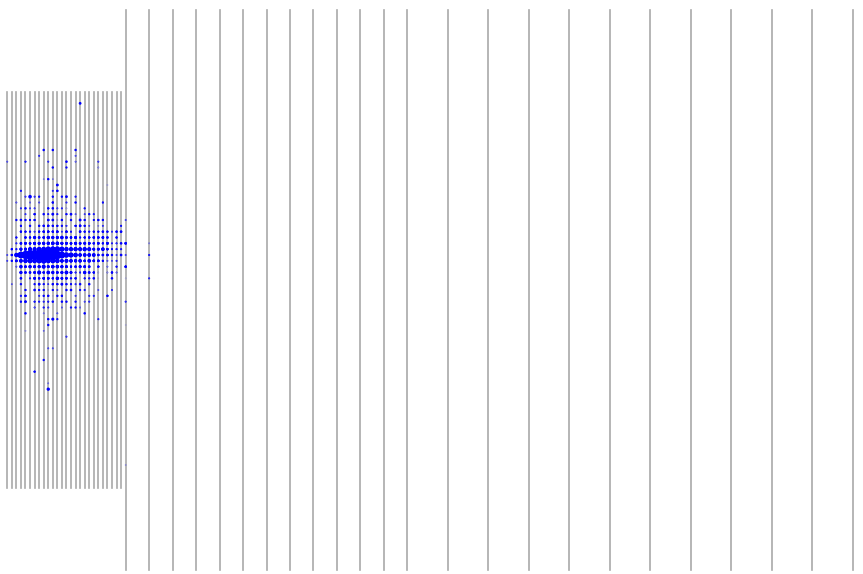}
				\end{subfigure}
				\hfill
				\begin{subfigure}{0.48\textwidth}
					\centering
					\includegraphics[width=\linewidth]{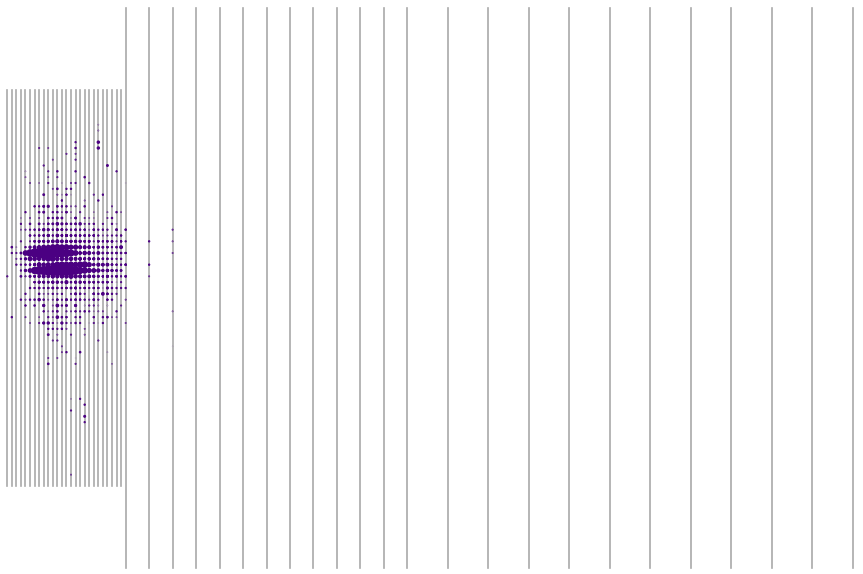}
				\end{subfigure}
				\hfill
				\begin{subfigure}{0.48\textwidth}
					\centering
					\includegraphics[width=\linewidth]{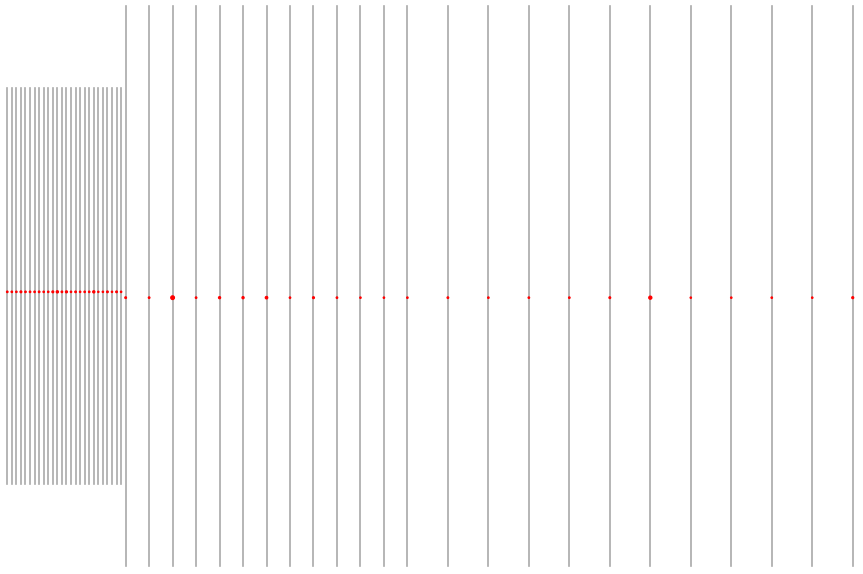}
				\end{subfigure}
				\hfill
				\begin{subfigure}{0.48\textwidth}
					\centering
					\includegraphics[width=\textwidth]{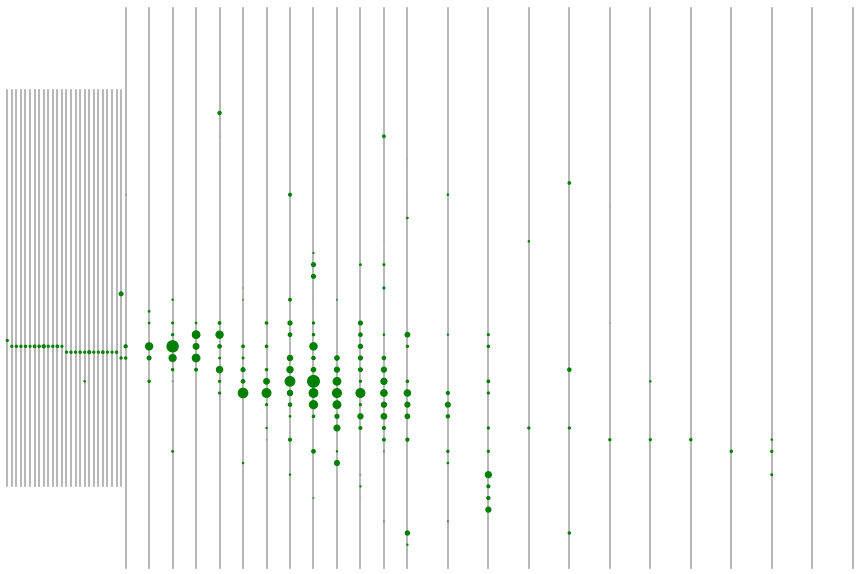}
				\end{subfigure}
				\caption{Typical signatures left by $e^-$ (blue), $\gamma\gamma$ (purple), $\mu$ (red) and $\pi^+$ (green) in the D2 detector. One clearly sees the dual shower core structure for $\gamma\gamma$ events.}\label{sec:d2-fig:event_displays}
			\end{figure}

			We observe the typical behaviours for these particles in this energy range in the detectors: $e^-/\gamma$s shower in the ECAl, $\mu$s traverse the full detector as MIPs, leaving a single ionisation track along its path, and $\pi^+$s leave a track in the ECAL and eventually shower in the HCAL. Figure \ref{sec:d2-fig:event_displays} shows typical event displays for particles entering the detector collinearly with the longitudinal axis.
			
			The dataset includes detailed informations on each particle event, namely the energy measurement in each cell but also extracted variables of interest describing the geometry and the energy repartition of the hits.
			
			From photon events, we also construct $\gamma$ pair events, by overlapping pairs of showers, with any cell hit by both showers taking as energy the sum of energy depositions from either showers. The motivation behind the creation of this data class is that neutral $\pi^0$ decay into a pair of photons, which then individually shower in the detector.
			
\begin{figure}
				\centering
				\includegraphics[width=0.5\textwidth]{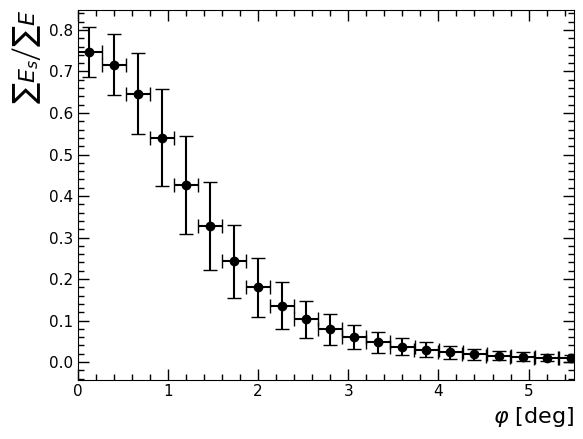}
				\caption{Ratios of the shared deposited energy and the total energy deposited energy by photons in the D2 detector, as a function of separation angle $\varphi$. A balance is struck by keeping $\varphi$ in $[1, 4.5]^\circ$.}\label{sec:d2-fig:shared_energy}
			\end{figure}

			When generating the overlapped shower set of events, one must strike a balance between ensuring that the showers are resolved enough to successfully treat them as a two $\gamma$ event, while avoiding the less interesting case of fully resolved showers. To strike this balance, we only consider $\gamma$s which form an angle $\varphi$ between their velocities such that $1^\circ \leq \varphi \leq 4.5^\circ$. This balance was struck by looking at the fraction of deposited energy that is shared amongst either showers over the full energy, shown in Figure \ref{sec:d2-fig:shared_energy}.
	
	\section{Optimised Graph Neural Network Operations}\label{sec:algos}
	
		\subsection{Graph Construction}
			
			Every detector event is represented as a graph, the nodes of which correspond to every hit of the event, linked by arbitrarily defined edges. This step is the main bottleneck of Graph-based pipelines, as graph construction algorithms, such as the $k$-nearest neighbours (KNN) algorithm, are usually quadratic with the number of nodes. To lower this prohibitive complexity for graph construction, our edge building strategy is based off of so-called Proximity Tables (PTs), which contain a pre-computed order of neighbourhood, for every sensor of the detector. That is, for every sensor of the detector, all other sensors are ordered according to a user-defined metric, as in Table \ref{sec:graph_algo-subsec:ptknn-tab:pt}. The graph building procedure then follows Algorithm \ref{sec:graph_algo-subsec:ptknn-algo:pt-knn}.
			
			\begin{table}
				\centering\
				\begin{tabular}{c||cccccc}
					Sensor ID & pos. 1   & pos.2    & $\cdots$ & pos. $j$ & $\cdots$ & pos. $M-1$  \\ \hline \hline
					1         & $s_1^1$  & $s_1^2$  & $\cdots$ & $s_1^j$  & $\cdots$ & $s_1^{M-1}$ \\
					2         & $s_2^1$  & $s_2^2$  & $\cdots$ & $s_2^j$  & $\cdots$ & $s_2^{M-1}$ \\
					$\vdots$  & $\vdots$ & $\vdots$ & $\ddots$ & $\vdots$ & $\ddots$ & $\vdots$    \\
					$i$       & $s_i^1$  & $s_i^2$  & $\cdots$ & $s_i^j$  & $\cdots$ & $s_i^{M-1}$ \\
					$\vdots$  & $\vdots$ & $\vdots$ & $\ddots$ & $\vdots$ & $\ddots$ & $\vdots$    \\
					$M$       & $s_M^1$  & $s_M^2$  & $\cdots$ & $s_M^j$  & $\cdots$ & $s_N^{M-1}$
				\end{tabular}
				\caption{Proximity table. $s_i$ designates the sensor ID'd $i$, $s_i^j$ is the sensor in $j$-th position in the row for $s_i$. $M$ is the number of sensors in the detector. Letting $m(s_i, s_i^j)$ be the value of the metric used to order the table, the neighbours of each sensors are ordered such that for any sensor $s^i$ and any $j$, $m(s_i, s_i^j) \leq m(s_i, s_i^{j+1})$.}\label{sec:graph_algo-subsec:ptknn-tab:pt}
			\end{table}
			
			\begin{algorithm}
				\caption{PTKNN: Proximity table $k$ nearest neighbours graph generation algorithm. $k$ is the number of neighbours wanted for each node.}\label{sec:graph_algo-subsec:ptknn-algo:pt-knn}
				\begin{algorithmic}[1]
					\State \textbf{Input}: hits, $k$
					\State \textbf{Output}: $G=\{V, E\}$
					\State $V \leftarrow \{\}$
					\State $E \leftarrow \{\}$
					\For{$s_i \in$ hits}
						\State $V \leftarrow V \cup \{s^i\}$
						\State counter $\leftarrow$ 0
						\While{counter $< k$}
							\For{$j: 1 \to M-1$}
								\If{$s_i^j \in$ hits}
									\State $V \leftarrow V \cup \{s_i^j\}$
									\State $E \leftarrow E \cup \{(s_i, s_i^j)\}$
									\State counter $\leftarrow$ counter$+ 1$
								\EndIf
							\EndFor
						\EndWhile
					\EndFor
					\State \textbf{return} $G = \{V, E\}$
				\end{algorithmic}
			\end{algorithm}
			
			The different metrics that can be defined can be used to enhance or limit some structural properties of the produced graph, such as for example the number of connected components or the assortativity coefficient (similarity of connections in the graph with respect to the node degree) of the graph. The simplest choice would be to use the Euclidean distance as a metric (making PT-KNN strictly equivalent to classical KNN), but one could think of metrics that contain some physical information, by using, for instance, the correlation of sensors as a metric, or by modifying the euclidean distance to limit the tendency of classical KNN to produce multiple connected component. An example is what we called the ``radiality distance'' metric, defined as
			\begin{equation}
				m(s_i, s_j) = (1+\alpha r) \Vert s_i - s_j \Vert,
			\end{equation}
			with $r$ the ``radiality factor'' of the sensor pair $(s_i, s_j)$, defined as the radial coordinate of the barycentre $m$ of $s_i$ and $s_j$, and $\alpha$ is the user-defined ``radiality coefficient'', modulating the importance of the radiality in the metric. It is also possible to use multiple PTs to build the graph, for example looking for $k_{\text{in}}$ neighbours in one PT of sensors in the same layer and $k_{\text{out}}$ neighbours in another PT of sensors in other layers.
			In order to choose the best metric, we trained different pipelines for each metric to compute the correlation of the accuracy of the trained model and of graph properties impacted by the metric choices\footnote{The selected metrics were : number of edges, average clustering, number of connected components, density, degree assortativity coefficient, global/local efficiency, diameter, computed using the NetworkX python package \cite{hagberg_08}.}. We however observed no strong dependence of both the classification and regression tasks on the metric choice. For simplicity, we only use, from now on the Euclidean distance for graph construction.
			
\begin{figure}
				\centering
				\includegraphics[width=0.5\textwidth]{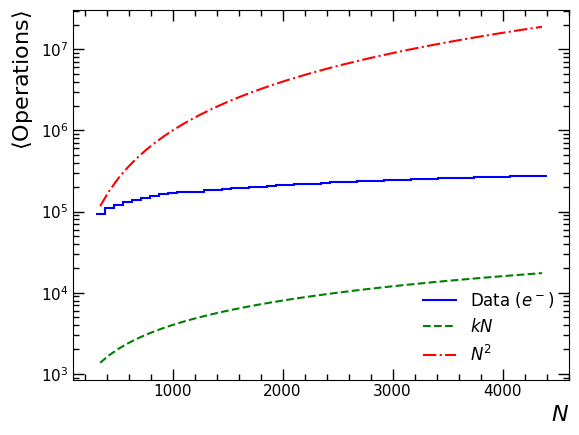}
				\caption{Average number of operations for the PT-KNN algorithm, using $k=4$ for $e^-$ events of the D2 dataset (see section \ref{chap:sim-sec:d2-subsec:events}). A comparison shows linear and quadratic complexities. Asymptotically, the average number of operations behaves like $\mathcal{O}(kN)$, with $N$ the number of nodes.}\label{chap:graph_algo-sec:ptknn-fig:pt_knn_complexity}
			\end{figure}

			Using PTs for KNN drastically reduces the complexity of the algorithm. Indeed, while the best case complexity of KNN is $\mathcal{O}\left(N^2\right)$ in the number of nodes $N$, the PT-KNN best case complexity is $\mathcal{O}(kN)$. In practice, if we look at the average last explored column $\left\langle c_\text{last} \right\rangle$ in the proximity table, we observe  a clear improvement from quadratic complexity, as shown in Figure \ref{chap:graph_algo-sec:ptknn-fig:pt_knn_complexity}.
			We look for the formula of the average last column $\left\langle c_\text{last} \right\rangle$ explored by the PT-KNN algorithm. We look at the central point of a Gaussian distribution of nodes centred at the origin, and assume that the statistics allow us to consider it as a general case.
			
			The probability to find a node at a distance $r$ from the origin is given by a half-Gaussian
			$$
			p(r) = \frac{\sqrt{2}}{\sigma \sqrt{\pi}} \exp \left( -\frac{r^2}{2 \sigma^2} \right),
			$$
			with $\sigma$ the standard deviation of distances from the origin. The number of cells at radius $r$ is $c(r) = ar$, with $a$ an arbitrary parameter, which can be verified with a square grid, for instance. Hence the number of nodes (or objects) at radius $r$ is given by
			$$
			o(r) = \mathcal{N} c(r) p(r),
			$$
			with $\mathcal{N}$ a renormalisation factor to ensure that $n$ nodes be found in $[0, \infty)$. The renormalisation factor is defined to ensure that 
			$$
			\int_0^\infty \mathcal{N} c(r) p(r) dr = N,
			$$
			and therefore 
			$$
			\mathcal{N} = \frac{N \sqrt{\pi}}{a \sqrt{2} \sigma} \quad \Rightarrow \quad o(r) = \frac{N}{\sigma^2} r \exp \left( -\frac{r^2}{2 \sigma^2} \right).
			$$
			We look for the radius $\tilde{r}$ such that $k$ neighbours have been found. That is, we solve for $\tilde{r}$
			$$
			\int_0^{\tilde{r}} \frac{N}{\sigma^2} r \exp \left( -\frac{r^2}{2 \sigma^2} \right) = k,
			$$
			giving $\tilde{r}^2 = 2 \sigma^2 \ln \left( \frac{N}{N-k} \right)$.
			Finally, we count the number of cells that were explored before reaching radius $\tilde{r}$, i.e. $\left\langle c_\text{last} \right\rangle$
			\begin{align*}
				\left\langle c_\text{last} \right\rangle & = \int_0^{\tilde{r}} ar dr \\
				& = a \sigma^2 \ln \left( \frac{N}{N-k} \right).
			\end{align*}
			Note however that for large $N$, this tends to zero, when we know that $\left\langle c_\text{last} \right\rangle \geq k$. This is because the previous integral was taken from $r=0$ when in truth there exists a minimal radius $r_\text{min}$ before one may find neighbours. To correct for this, we force a lower bound $k$ by adding a term 
			$$
			\left\langle c_\text{last} \right\rangle = a \sigma^2 \ln \left( \frac{N}{N-k} \right) + k.
			$$
			One last remark that must be made is that because we have a half-Gaussian PDF, the parameter $\sigma$ is not the standard deviation of the distances from the centre to the nodes, which is related to the true standard deviation $s$ of the node distances to the origin as
			$$
			\sigma = s\sqrt{\frac{\pi}{\pi-2}}.
			$$
			\begin{figure}
				\centering
				\includegraphics[width=0.5\textwidth]{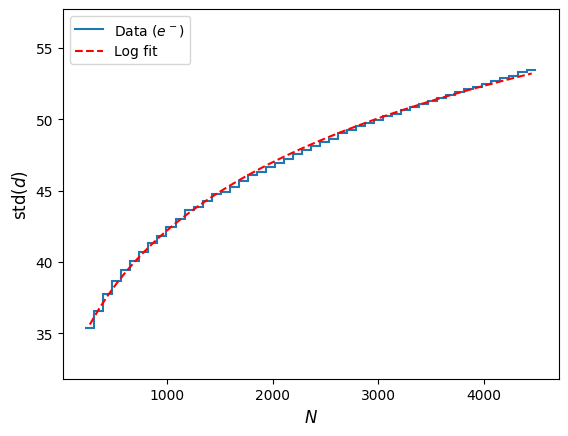}
				\caption{Standard deviation of the distance to the origin of nodes. This data is obtained from centred $e^-$ events of the D2 dataset, using $k=4$. $N$ is the bumber of nodes, and $d$ is the distance between the nodes and the origin.}\label{chap:graph_algo-sec:pt-fig:fit_sigma}
			\end{figure}
			Fitting the standard deviation on the data with a logarithm (the number of available cells as we increase the radius increases exponentially, see Figure \ref{chap:graph_algo-sec:pt-fig:fit_sigma}), we find that the average last selected column should be
			$$
			\left\langle c_\text{last} \right\rangle = A \ln^2(N) \ln \left( \frac{N}{N-k} \right) + k,
			$$
			meaning that $\left\langle c_\text{last} \right\rangle = \mathcal{O} \left( \frac{\log^2(N)}{N} \right)$, which is indeed verified in the relevant node count range ($\sim 1000$). If we look at infinity, this means that the number of operations reduces to $k$. This is verified, since at infinity, every single cell should be filled, meaning that all neighbours can be found in the closest geometric neighbours of each cell.
			
\begin{figure}
				\centering
				\begin{subfigure}{0.49\textwidth}
					\centering
					\includegraphics[width=\textwidth]{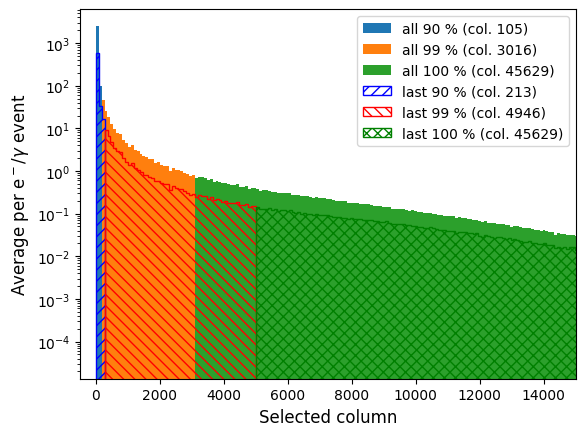}
				\end{subfigure}
				\begin{subfigure}{0.49\textwidth}
					\centering
					\includegraphics[width=\textwidth]{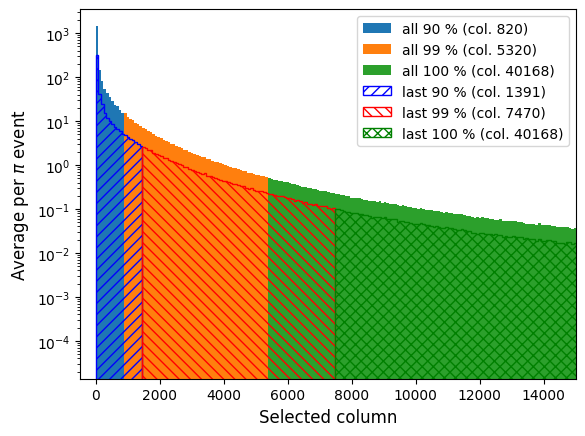}
				\end{subfigure}
				\caption{Distribution of column indices for node selection using the PT-KNN algorithm for $e^-$ and $\gamma$ events (left), and $\pi^+$ events (right), using $k=4$.}\label{chap:graph_algo-sec:pt-fig:last_col_distrib}
			\end{figure}

			We can further reduce the mean complexity of the PT-KNN algorithm by truncating the proximity table to a given number of columns $c_\text{max}$. In doing so, we reduce the worst-case complexity of the PT-KNN algorithm to $\mathcal{O}(c_\text{max} N)$, which, if we are aggressive enough with the PT truncation, can go below the average number of operations. To see to what extent this truncation is doable, we look at the distribution of columns which contained neighbours when running the PT-KNN algorithm for electron and pion events in D2 respectively. This is shown in Figure \ref{chap:graph_algo-sec:pt-fig:last_col_distrib}.
			One observes that the columns in which neighbours are selected are quite close to the beginning, and that most of the edges are built by exploring the first few hundred columns. We observe in particular that if we truncate the PT at 5320, we still build 99 \% of all edges, the remaining 1 \% being for outlier nodes that appear far away from the shower core. While removing the last 1 \% will have a minor impact on the average number of operations, it will ensure a lower upper bound for the number operations, essential in highly constrained contexts like particle detector data analysis \cite{goldreich_11}. Note that from the high connectivity ensured by relatively high values of $k$ and undirected graphs, we can actually aggressively truncate the PT with a limited impact on information flow through the graph. We therefore can reduce the average complexity with a limited impact on regression performance. Since the edges that are lost are long edges, this will in fact have a tendency to smoothen the graphs' variances, ensuring a stringer connectivity close to the shower core.

		\subsection{Grpah Convolution Operation}
			
			The core operation of the modern implementation of Graph neural networks is the graph convolution operation. In this work, we use the Message passing convolution operation, defined by Gilmer et al. \cite{gilmer_17}: given a graph $G = \{V, E\}$, the message passing convolution operation updates the features of each node $v\in V$ from $\zeta_v$ to $\xi_v$ as
			\begin{equation}\label{eq:mp_generic_eq}
				\xi_v = \gamma_{\Theta_\gamma} \left( \zeta_v, \underset{v\in \mathcal{N}(v)}{\square} \phi_{\Theta_\phi} \left(\zeta_v, \zeta_w, e_{vw} \right) \right),
			\end{equation}
			where $\phi_{\Theta_\phi}$ is the message function, combining the features of $v$, of a neighbour $w\in \mathcal{N}(v)$ of $v$, and the features of their linking edge $e_{vw}\in E$. It is parametrised by trainable weights $\Theta_\phi$, an is usually implemented as a shallow neural network. $\square$ is the aggregator, updating the current node's features with the aggregated messages from neighbours. Since the number of neighbours of $v$ may vary, the aggregation operator must be normalised and commutative. Typical aggregators are max or mean functions. $\gamma_{\Theta_\gamma}$ is the update function, mixing the aggregated messages and the node features. It can be very simple or implemented as a shallow MLP, and has trainable parameters $\Theta_\gamma$.
			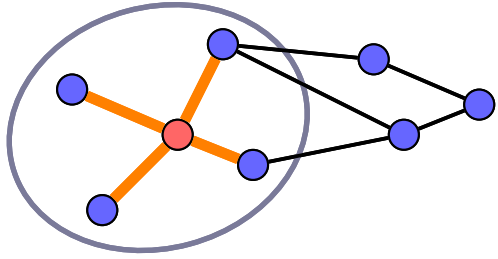
\begin{figure}
				\centering
				\resizebox{0.4\columnwidth}{!}{
					\begin{tikzpicture}
						\node[circle, draw, fill=red!60, inner sep=-2] (v1) at (-0.3, 0.2) {};
						\node[circle, draw, fill=blue!60, inner sep=-2] (v2) at (-1, 0.5) {};
						\node[circle, draw, fill=blue!60, inner sep=-2] (v3) at (-0.8, -0.3) {};
						\node[circle, draw, fill=blue!60, inner sep=-2] (v4) at (0, 0.8) {};
						\node[circle, draw, fill=blue!60, inner sep=-2] (v5) at (0.2, 0) {};
						\draw[-, line width=2pt, color=orange] (v1) -- (v2);
						\draw[-, line width=2pt, color=orange] (v1) -- (v3);
						\draw[-, line width=2pt, color=orange] (v1) -- (v4);
						\draw[-, line width=2pt, color=orange] (v1) -- (v5);
						\draw[-, line width=1pt, rotate=15, color=blue!20!white!60!black] (-0.35,0.35) ellipse (1 and 0.8);
						
						\node[circle, draw, fill=blue!60, inner sep=-2] (v6) at (1, 0.7) {};
						\node[circle, draw, fill=blue!60, inner sep=-2] (v7) at (1.2, 0.2) {};
						\node[circle, draw, fill=blue!60, inner sep=-2] (v8) at (1.7, 0.4) {};
						\draw[-, line width=0.7pt] (v4) -- (v6);
						\draw[-, line width=0.7pt] (v4) -- (v7);
						\draw[-, line width=0.7pt] (v5) -- (v7);
						\draw[-, line width=0.7pt] (v6) -- (v8);
						\draw[-, line width=0.7pt] (v7) -- (v8);\node[circle, draw, fill=red!60, inner sep=-2] (v1) at (-0.3, 0.2) {};
						\node[circle, draw, fill=blue!60, inner sep=-2] (v2) at (-1, 0.5) {};
						\node[circle, draw, fill=blue!60, inner sep=-2] (v3) at (-0.8, -0.3) {};
						\node[circle, draw, fill=blue!60, inner sep=-2] (v4) at (0, 0.8) {};
						\node[circle, draw, fill=blue!60, inner sep=-2] (v5) at (0.2, 0) {};
						\draw[-, line width=2pt, color=orange] (v1) -- (v2);
						\draw[-, line width=2pt, color=orange] (v1) -- (v3);
						\draw[-, line width=2pt, color=orange] (v1) -- (v4);
						\draw[-, line width=2pt, color=orange] (v1) -- (v5);
						\draw[-, line width=1pt, rotate=15, color=blue!20!white!60!black] (-0.35,0.35) ellipse (1 and 0.8);
						
						\node[circle, draw, fill=blue!60, inner sep=-2] (v6) at (1, 0.7) {};
						\node[circle, draw, fill=blue!60, inner sep=-2] (v7) at (1.2, 0.2) {};
						\node[circle, draw, fill=blue!60, inner sep=-2] (v8) at (1.7, 0.4) {};
						\draw[-, line width=0.7pt] (v4) -- (v6);
						\draw[-, line width=0.7pt] (v4) -- (v7);
						\draw[-, line width=0.7pt] (v5) -- (v7);
						\draw[-, line width=0.7pt] (v6) -- (v8);
						\draw[-, line width=0.7pt] (v7) -- (v8);
					\end{tikzpicture}
				}
				\caption{Illustration of the graph convolution operation. Nodes (in red) collect information from their neighbours as the ``messages'' (orange), and use the aggregation of all messages to update their own features.}
			\end{figure}
			The idea of graph convolution is that each node collects  a collection of messages from all of its neighbours. Once all messages are collected, the aggregator makes a synthesis of the messages. This synthesis of its surroundings is what is used by the node to update its features.
			
			Our implementation simplifies the message passing operation by adding self-loops, , i.e. considering $v$ as one of its own neighbours. Doing this is equivalent to implementing an update function $\gamma_{\Theta_\gamma}$ \cite{gilmer_17, kipf_17}. The message is computed by concatenating the features $\zeta_v, \zeta_{w\in \mathcal{N}(v)}$ and $e_{vw}\in E$, and applying a shallow neural network on it. The resulting function is 
			\begin{equation}\label{eq:mp_ours}
				\xi_v = \underset{v\in \tilde{\mathcal{N}}(v)}{\square} \phi_{\Theta} \left( \text{concat} \left[\zeta_v, \zeta_w, e_{vw} \right] \right),
			\end{equation}
			where $\tilde{\mathcal{N}}(v) = \mathcal{N}(v) \cup \{v\}$, and $\phi_\Theta$ is a neural network with weights $\Theta$. The length of self loop edges is set to $e_{vv}=0$.
		
		\subsection{Graph Pooling}\label{sec:treclus}
			
			To increase the receptive field of the convolution operation, i.e. increasing the range of messages, it is usual to pool the graph, reducing the node count to look at the graph at a broader scale. We study graph pooling as presented by Grattarola et al. \cite{grattarola_24}, which divides the pooling operation in three steps: a clustering (or selection) step, where we select which nodes will be clustered together, followed by a reduction step, where nodes within clusters are pooled, and finally a connection step, during which we update the adjacency of clusters.
			
\begin{figure}
				\centering
				\includegraphics[width=0.9\textwidth]{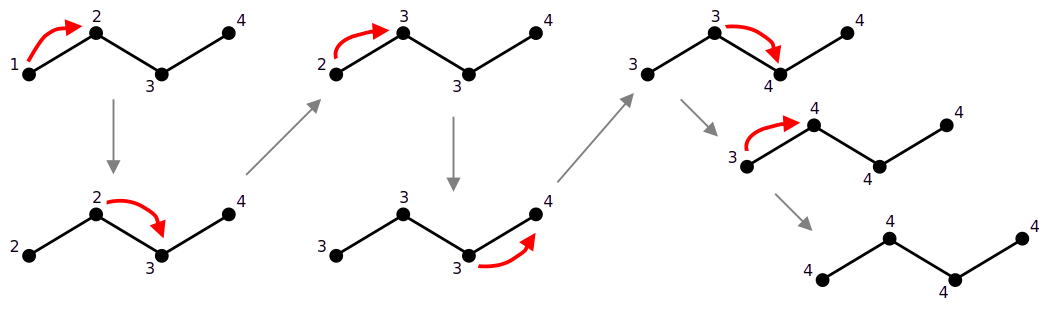}
				\caption{Chain cluster, where clustering edges share end vertices. Unfortunate winner assignment might make the clustering complexity explode.}\label{sec:treclus-fig:chain_cluster}
			\end{figure}

			The graphs are embedded in a euclidean space, and we therefore want to cluster together neighbours that are close to one another, i.e. make single nodes from node neighbourhoods. To this end, we developed a threshold clustering algorithm, called Treclus, that pairs neighbours that are close together, with their linking edge shorter than a threshold $\varepsilon$. To do so, it effectively creates a graph matching, which is defined for a graph $G=(V, E)$ as a subset of edges $E_m \subseteq E$ such that no two edges in $E_m$ share an end node \cite{lovasz_09}. Here, all edges in $E_m$ are shorter than the threshold 
			$\varepsilon$. It is very important that the clustering edges make a matching to keep a linear complexity of the clustering algorithm. Indeed, let us consider the cluster illustrated in Figure \ref{sec:treclus-fig:chain_cluster}.
			Assuming that clusters are labelled as the highest node label in their cluster. For the sake of simplicity, let us assume that the cluster belonging labelling procedure labels nodes 1 to 4. Node 1 gets assigned cluster label 2. Then node 2 gets cluster label 3, but all nodes labelled as cluster 2 must be relabelled to cluster 3 as well. Finally, node 3 is labelled as cluster 4, and we must update the cluster labels of nodes 1 and 2. Any order of cluster labelling results in a similar case of non-linear complexity, and any attempt at optimising the labelling order looses generality and symmetry. This problem is however avoided if clusters can only contain two nodes, ensuring that clustering from a matching has a linear complexity. While we ideally wish to find a maximum matching of the subgraph $G'=(V, E' = \{ e\in E \vert e \leq \varepsilon \})$, that is the matching that maximises the number of matching edges, maximal graph matching is usually a quadratic problem \cite{behnezhad_24}. Our implementation of the matching is therefore a brute force linear algorithm, as shown in \ref{sec:treclus-algo:matching}.
			
			\begin{algorithm}
				\caption{Treclus algorithm: Brute-force graph matching of graphs with thresholded edges. The returned tensor $C$ indexes each vertex to its pooling cluster. If clusters are reduced and connected, Treclus can be called multiple times to increase clustered edges.}\label{sec:treclus-algo:matching}
				\begin{algorithmic}[1]
					\State \textbf{Input}: $G=\{V, E\}$, $\varepsilon$
					\State \textbf{Output}: $C\in \mathbb{N}^N$
					\State $C \leftarrow [1, 2, \dots N = \vert V \vert]$
					\State $E_\varepsilon \leftarrow \{e_{vw} \in E \vert x_{vw} \leq \varepsilon\}$
					\State memory $\leftarrow \{\}$
					\For{$e_{vw} \in E_\varepsilon$}
						\If{$v, w \notin$ memory}
							\State winner, loser $\leftarrow$ random$\_$permutation$(v, w)$
							\State $C[$loser$] \leftarrow$ winner
							\State memory $\leftarrow$ memory$\cup \{v, w\}$
						\EndIf
					\EndFor
					\State \textbf{return} $C$
				\end{algorithmic}
			\end{algorithm}
			
\begin{figure}
				\centering
				\includegraphics[width=0.5\textwidth]{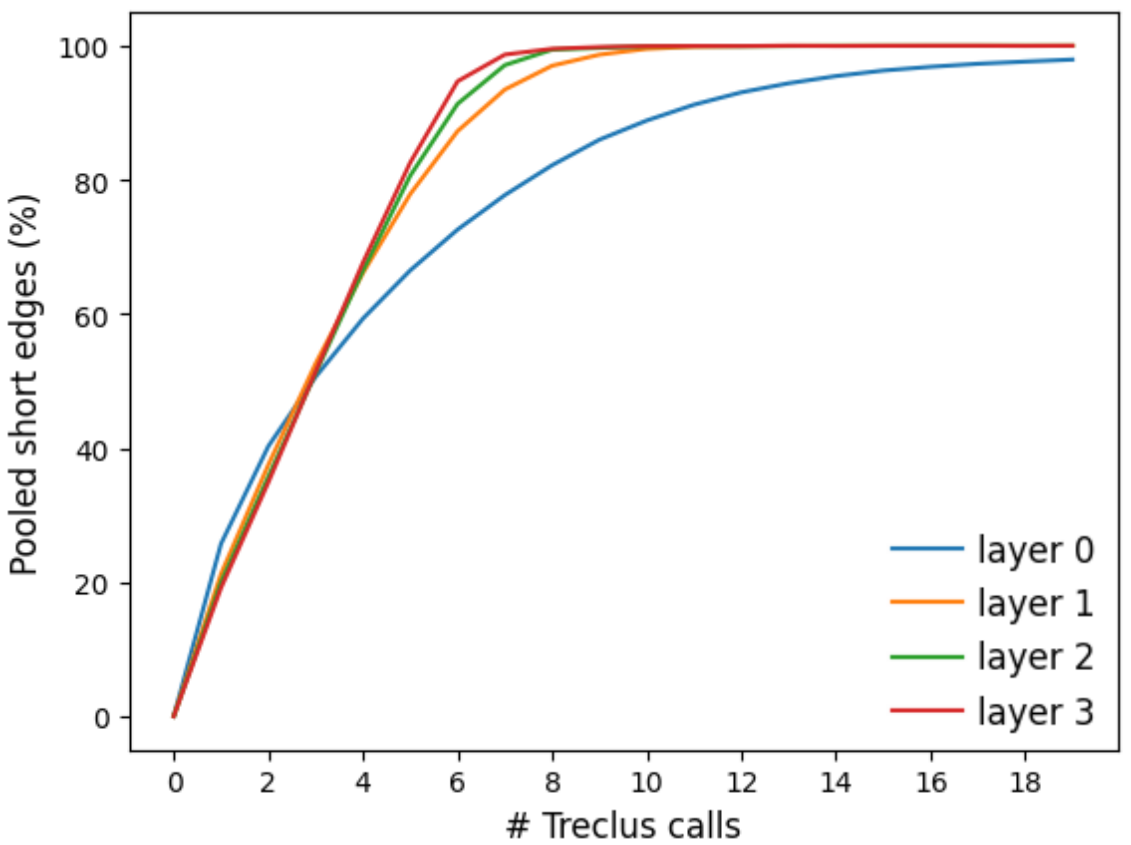}
				\caption{Fraction of pooled edges as a function of the number of Treclus calls. The study was done for $e^-$ events in D2, using as threshold the 30-th percentile of edge lengths.}\label{sec:treclus-fig:n_treclus}
			\end{figure}

			Note that the incomplete matching shortcoming can be solved by successive runs of the Treclus algorithm. Indeed, by deleting the edges between pairs of nodes clustered together, we may ensure that the repeated clustering performs a unique matching at every repeated step. We call this repeated Treclus clustering $N$-Treclus. where $N$ is the number of clustering runs. If we look at the proportion of edges shorter than $\varepsilon$ when running $N$-Treclus, we observe that most of the edges get clustered within the first iterations, as shown in Figure \ref{sec:treclus-fig:n_treclus}, computed using $e^-$ data points from the D2 dataset.
			Observe that the fraction of unpooled short edges decreases with the depth of the model and number of $N$-Treclus poolings. In particular, here we used as threshold the 30-th percentile of edge lengths at that layer, such that the final graph would have 10 to 50 nodes on average. This implies, first of all, that the number of edges decreases exponentially with the number of convolution layers, but also that any edge not clustered at layer $l$ may still be clustered at layer $l+1$. As a result, the number of nodes per cluster decreases from $\sim 10$ nodes to $<5$, coinciding with the number of needed Treclus calls shown in figure \ref{sec:treclus-fig:n_treclus}.
			
			One must also note that even a maximal matching cannot possibly cover all short edges, and that as such, even under the maximally matching assumption, multiple calls are needed. In any case, an order of less than 10 calls to a linear algorithm still beats quadratic complexity clustering algorithms. The pooling threshold also has a major impact on the number of required Treclus calls. Indeed, a high threshold will cover more edges, hence slowing down the clustering step. On the contrary, lowering $\varepsilon$ will lower the number of required operations. This can be kept in mind, and lower the number of required calls by imposing small thresholds in the first pooling layers, and gradually increasing it through the pipeline.
			
			The reduction step consists of a feature wise pooling of nodes within the same clusters. The choice of the pooling operation follows the same logic as the aggregator function choice for the convolution step. However, while this applies strictly for the classification task (with a max pooling), for the energy regression step, the aggregator must be a feature-wise sum. Indeed, if a cluster contained (without loss of generality) 2 nodes with identical features $x$ and another cluster contained 10 nodes with equal features $x$, using a mean pooling would imply that both resulting nodes have features $x$. We however want to keep the information that one cluster contained 2 nodes while the other contained 10, else adding or removing pooled nodes would not have any impact despite resulting from particles with different energies. A feature-wise sum would ensure that the factor 5 in node count of both clusters be significant.
			\begin{figure}
				\centering
				\includegraphics[width=0.65\textwidth]{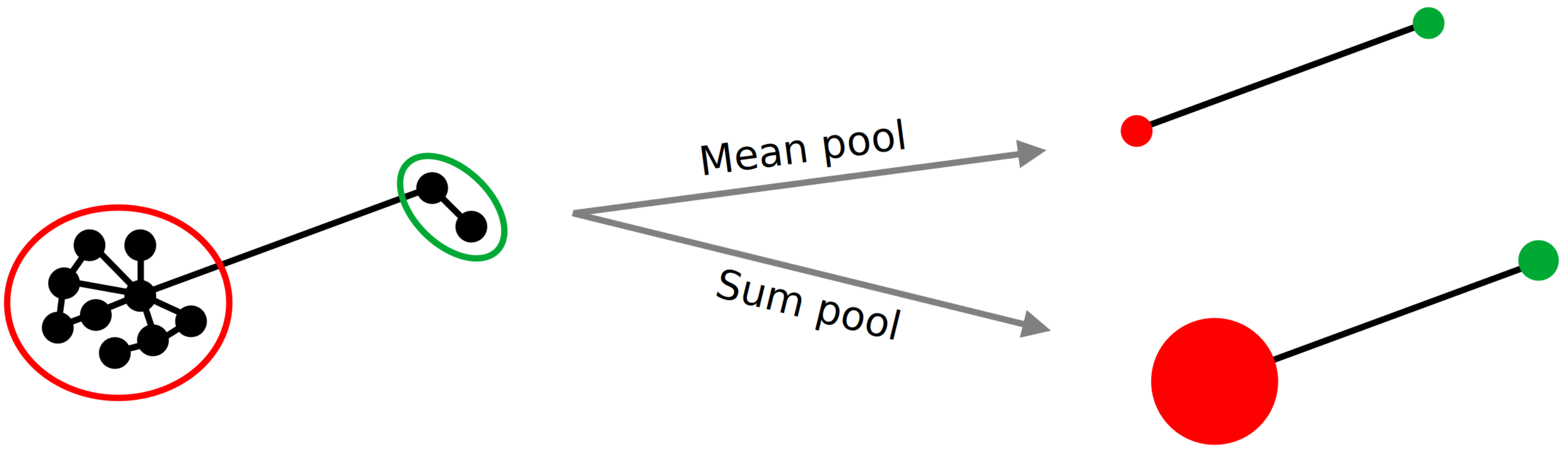}
				\caption{Using a sum aggregator for the reduction of large clusters ensure that node density information is maintained by the aggregator.}
			\end{figure}
			This might be surprising in comparison to image convolution, but one must keep in mind that the number of pixels pooled together is constant in CNNs. It is interesting to note however that different energies will result in different shower shapes. The Treclus algorithm keeping a good degree of the broad shape of a graph, the GNN is able to regress the energies with only a slight loss in performance, as GNNs are indeed very strong to infer information from the graph shape. The position of the cluster is taken at random from one of the nodes it contains. This allows to keep the graph embedded in the known detector geometry, avoiding inefficient operations further down the pipeline (see section \ref{sec:readout}), while preserving the locality of the cluster, since the length of the remaining edges makes the induced error negligible. The random choice in position however imposes an upper-bound on the threshold $\varepsilon$, which in turn limits the output size of the convolution operation to avoid overfitting, reducing the size of the network.
			
			Finally, for the reconnection step, the clusters inherit their adjacencies from the nodes that compose them, i.e. two clusters are neighbours if they each contain neighbouring nodes. The length of these edges is then re-computed to smoothen the accumulation of noise due to the random choice of cluster positions.
			
\begin{figure}
				\centering
				\begin{subfigure}{0.48\textwidth}
					\centering
					\includegraphics[width=\textwidth]{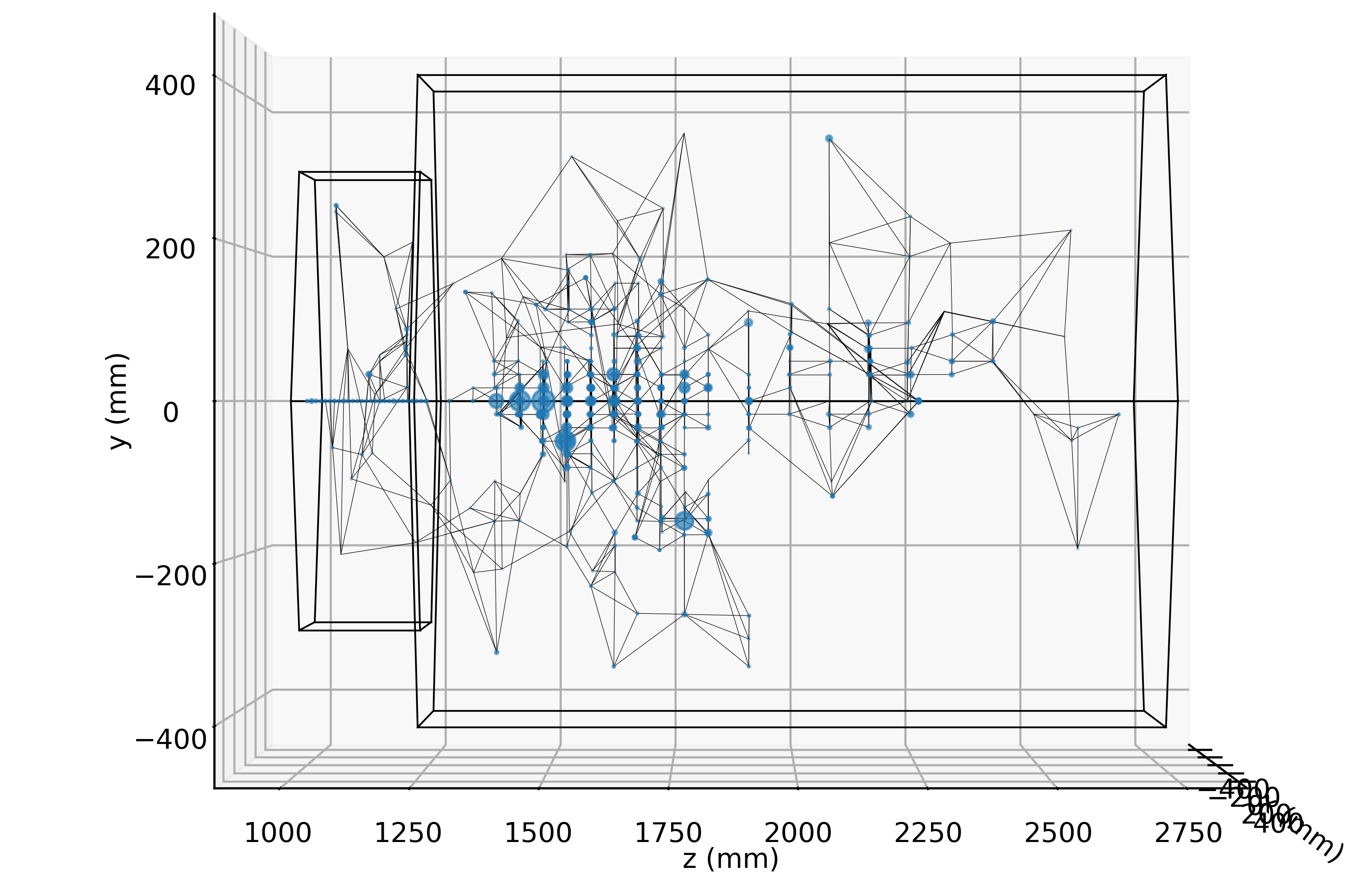}
				\end{subfigure}
				\begin{subfigure}{0.48\textwidth}
					\centering
					\includegraphics[width=\textwidth]{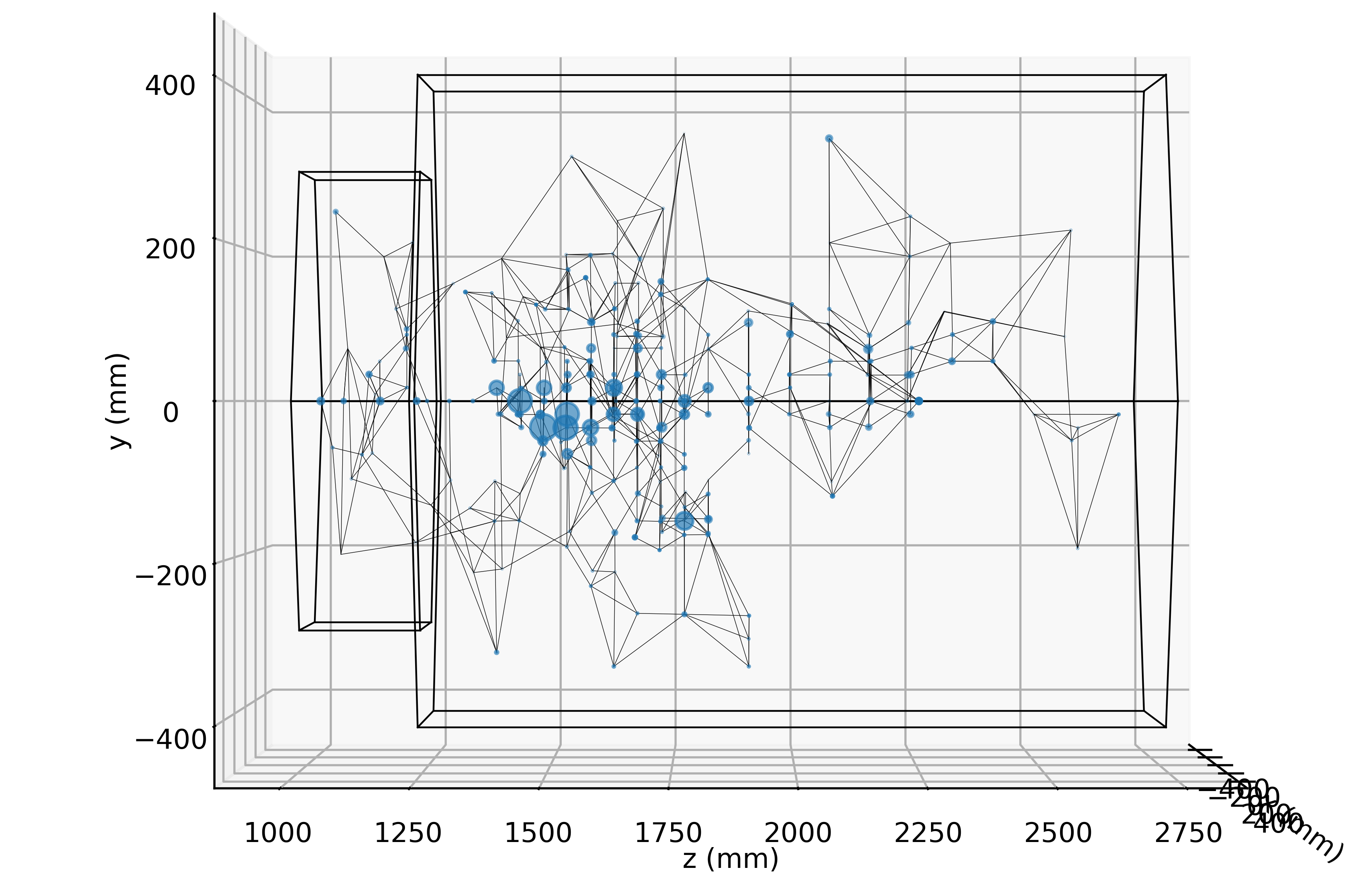}
				\end{subfigure}
				\begin{subfigure}{0.48\textwidth}
					\centering
					\includegraphics[width=\textwidth]{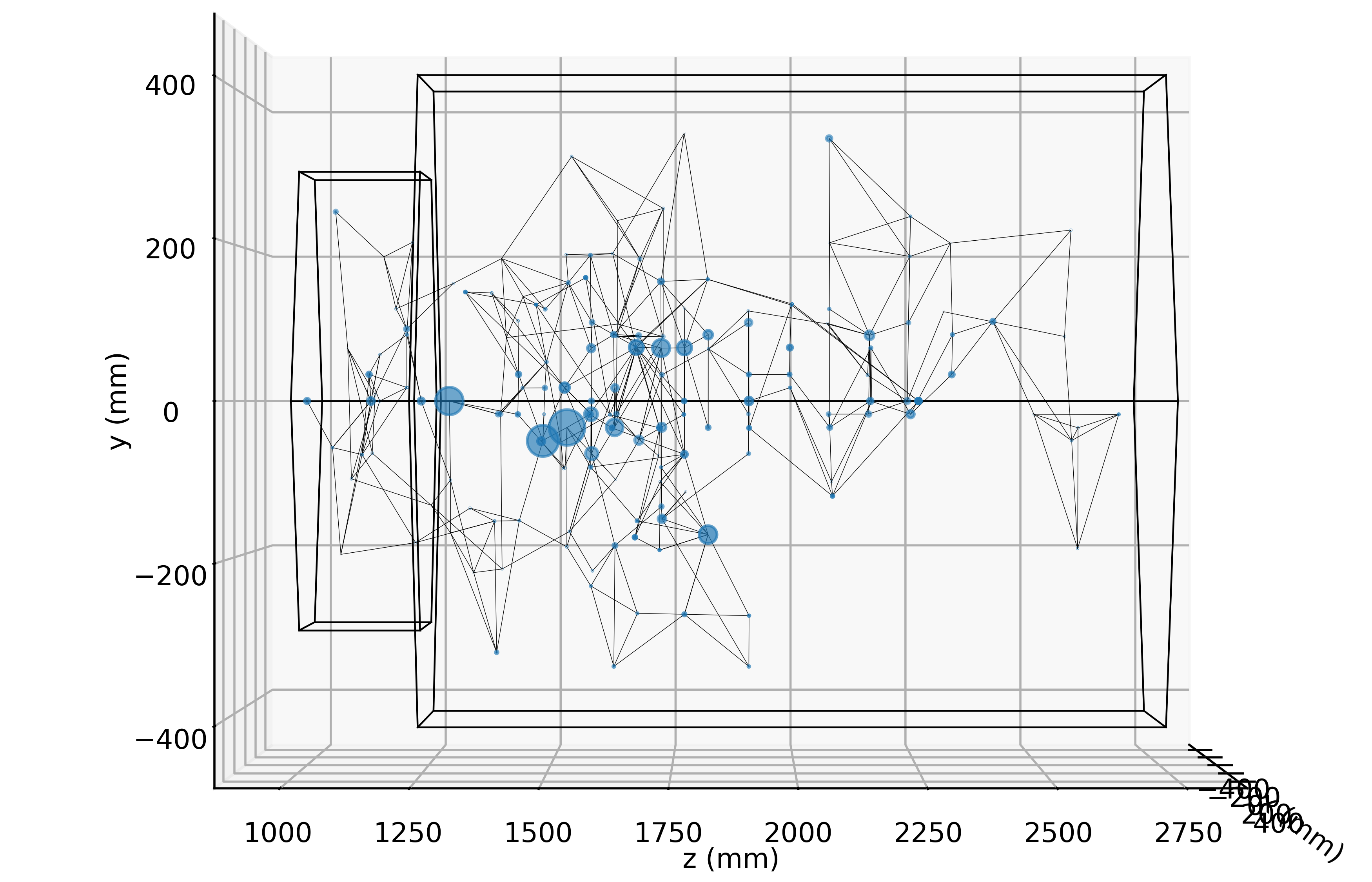}
				\end{subfigure}
				\begin{subfigure}{0.48\textwidth}
					\centering
					\includegraphics[width=\textwidth]{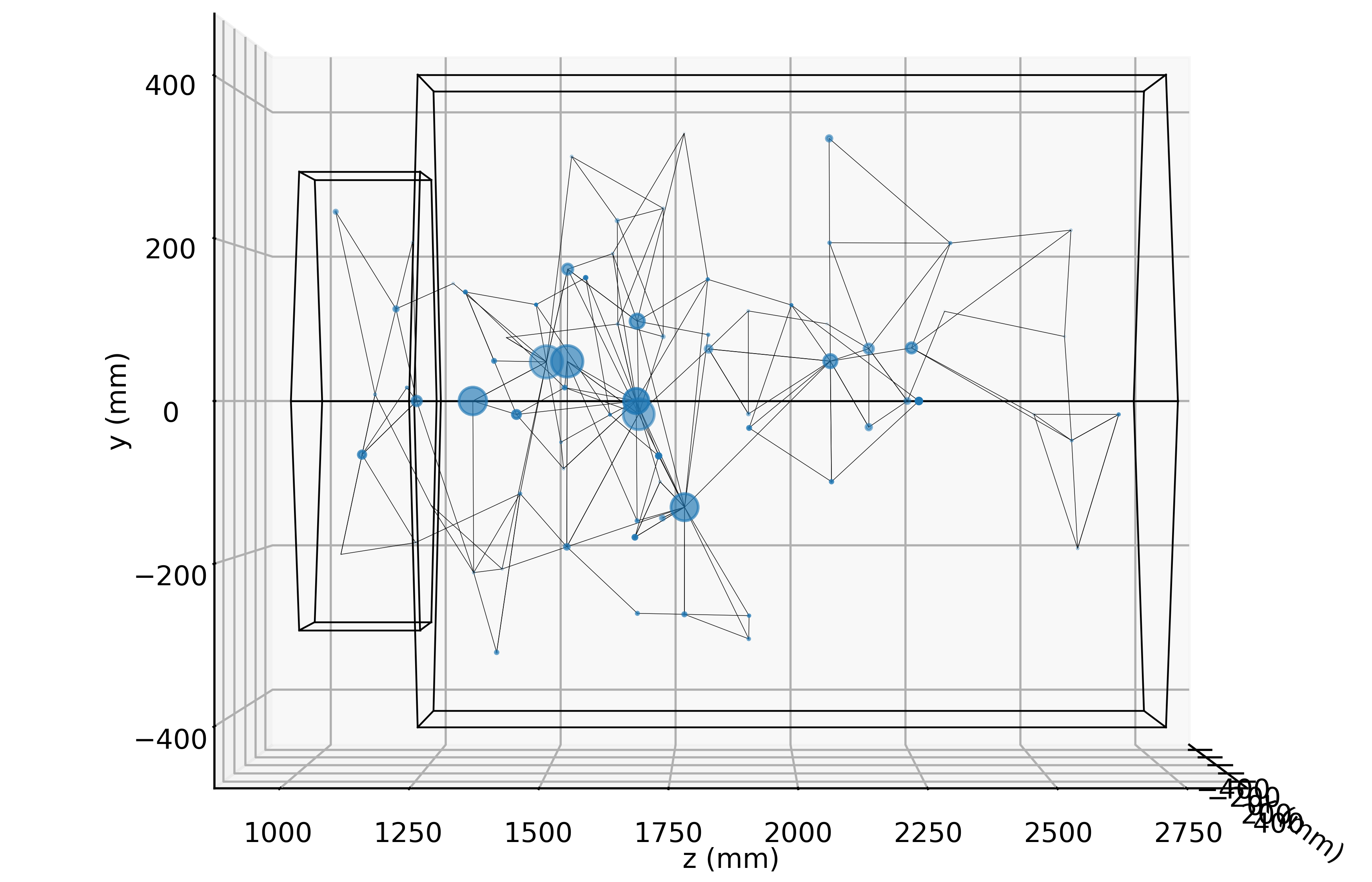}
				\end{subfigure}
				\caption{$\pi$ jet graph representation after successive pooling steps. We observe that while the number of nodes decreases, the global shape of the graph is preserved through pooling.}\label{sec:treclus-fig:full_calls}
			\end{figure}

			Looking at a typical graph through multiple layers of $N$-Treclus calls, as shown in Figure \ref{sec:treclus-fig:full_calls}, we observe that our objective of reducing the node density to increase the message passing convolution range, while preserving the overall graph structure, is successfully achieved.
			Note that if we are too restrictive on the number of columns $c_\text{max}$ in the proximity table, this pooling could result in disconnected components as it only maintains long edges, created by further explorations of the proximity table.
			
		\subsection{Readout Operation} \label{sec:readout}
			After the convolution and pooling layers, it is necessary to flatten the graph representation if the output of convolution needs to be fed into a Multi-layer perceptron (MLP). This operation, known as readout (or readout pooling), can be quite tricky, as graph-like data usually does not have any consistent ordering and may have variable sizes \cite{zhang_18_gnn_archi}. 
			
\begin{figure}
				\centering
				\begin{subfigure}{0.25\textwidth}
					\centering
					\includegraphics[width=\textwidth]{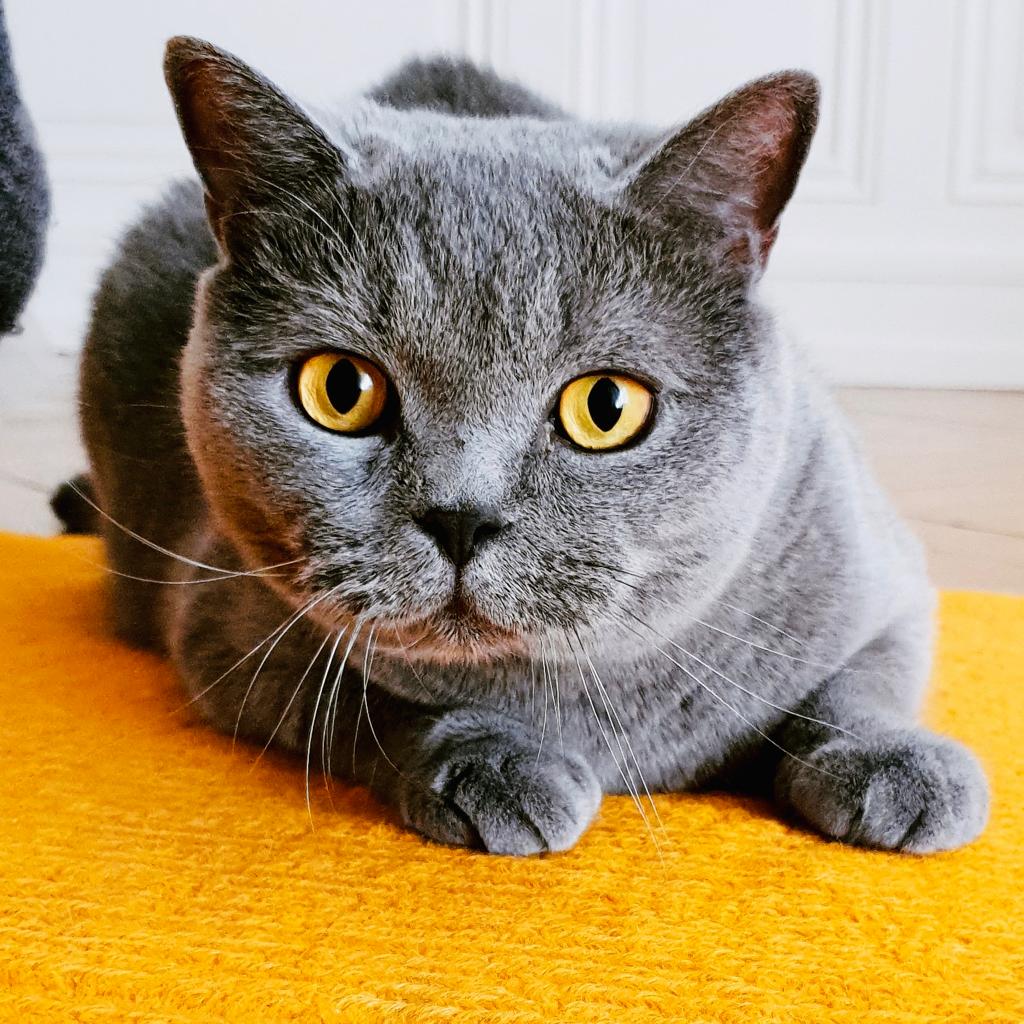}
					\caption{Original image}\label{sec:readout-fig:milo_shuffle-subfig:og_milo}
				\end{subfigure}
				\hspace{0.5cm}
				\begin{subfigure}{0.25\textwidth}
					\centering
					\includegraphics[width=\textwidth]{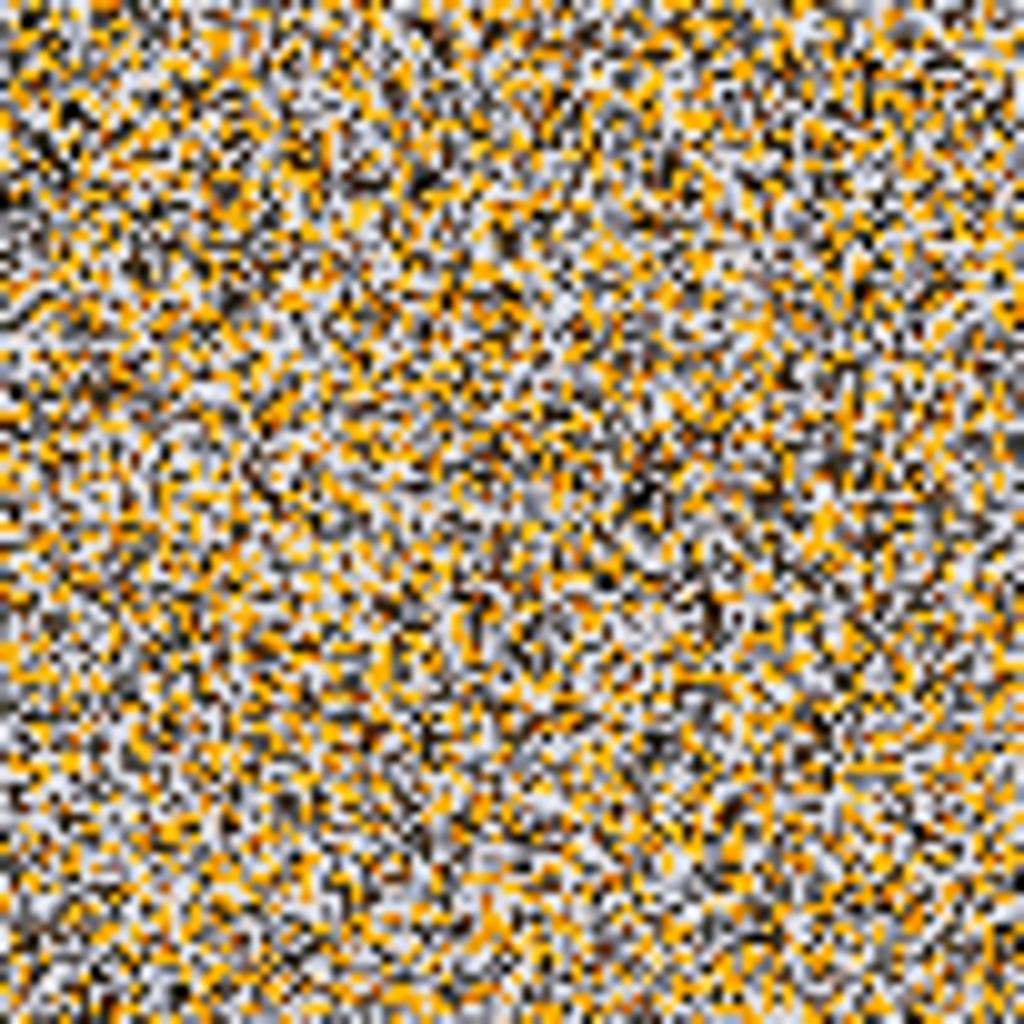}
					\caption{Shuffled image}\label{sec:readout-fig:milo_shuffle-subfig:mix_milo}
				\end{subfigure}
				\caption{The effect of a random readout on structural information loss.}\label{sec:readout-fig:milo_shuffle}
			\end{figure}

			The importance of ordering for the readout operation is illustrated in Figure \ref{sec:readout-fig:milo_shuffle}: looking at the original image \ref{sec:readout-fig:milo_shuffle-subfig:og_milo}, our brain interprets all pixels in a structured way and recognises the author's cat. Should the structure be random, as in \ref{sec:readout-fig:milo_shuffle-subfig:mix_milo}, it becomes impossible to interpret.
			The same applies to the readout of graphs and the ordering of nodes \cite{zhang_18_gnn_archi}.
			
			Our approach uses the fact that we know the detector geometry, and kept the nodes embedded in that geometry (see the reduction paragraph of Sec. \ref{sec:treclus}) to re-embed the graph back into the detector, and divide the detector into different Readout regions. This allows in particular to reduction in complexity by having a pre-computed mapping of cells to a readout region, not needing the systematic computation of node position belonging to readout regions. Using the same argument and procedure as in section \ref{sec:treclus}, all nodes within the same readout region are feature-wise pooled together, and the obtained feature of every region are flattened following a known order. It is however important that this voxelisation of the detector respects the geometry of the detector.
			
			For instance, in the case of the D2 dataset, we must respect the rotational symmetry of particle decays in the detector. If we rotated an electron shower around the longitudinal axis, while it is the same input, if the readout does not respect that symmetry in the detector, the model would have to learn both showers as two different inputs with identical output. This would hinder the training time and performance greatly. As such, we design the voxelisation procedure of the detector following its symmetries. In the case of the D2 detector, we slice the detector longitudinally, and subdivide each slice in concentric regions. The fact that the regions be concentric ensures the respect of rotational symmetry.
			\begin{figure}
				\centering
				\includegraphics[width=0.3\textwidth]{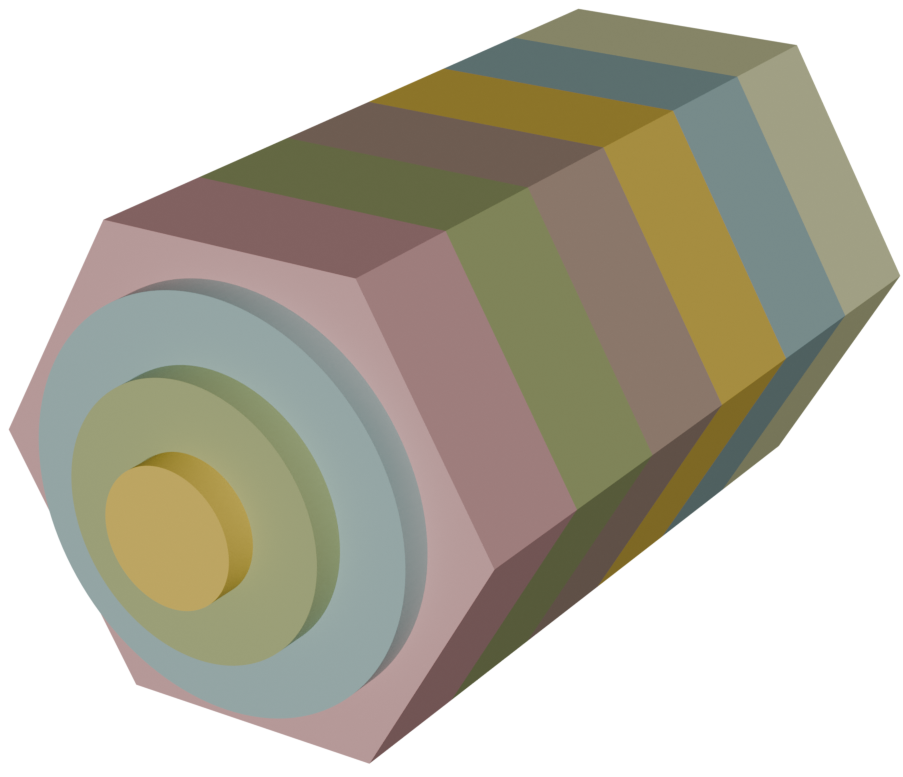}
				\caption{Radial readout regions of the detector. Each longitudinal slice of the detector is divided in concentric cylindrical regions, respecting radial symmetry. The design of the readout regions respects the rotational symmetry of the detector.}\label{sec:readout-fig:readout_regions}
			\end{figure}
			
			Note that the detector being composed of two different types of detector, electromagnetic and hadronic calorimeters, we can adapt the granularity of the readout to the task at hand: in energy regression tasks for electrons or photons, where there is no interaction with the hadronic part of the detector, a higher granularity of the readout is only needed in the ECAL.
			
		\subsection{Optimised GNN Pipelines}
		
\begin{figure}
				\centering
				\includegraphics[width=0.75\textwidth]{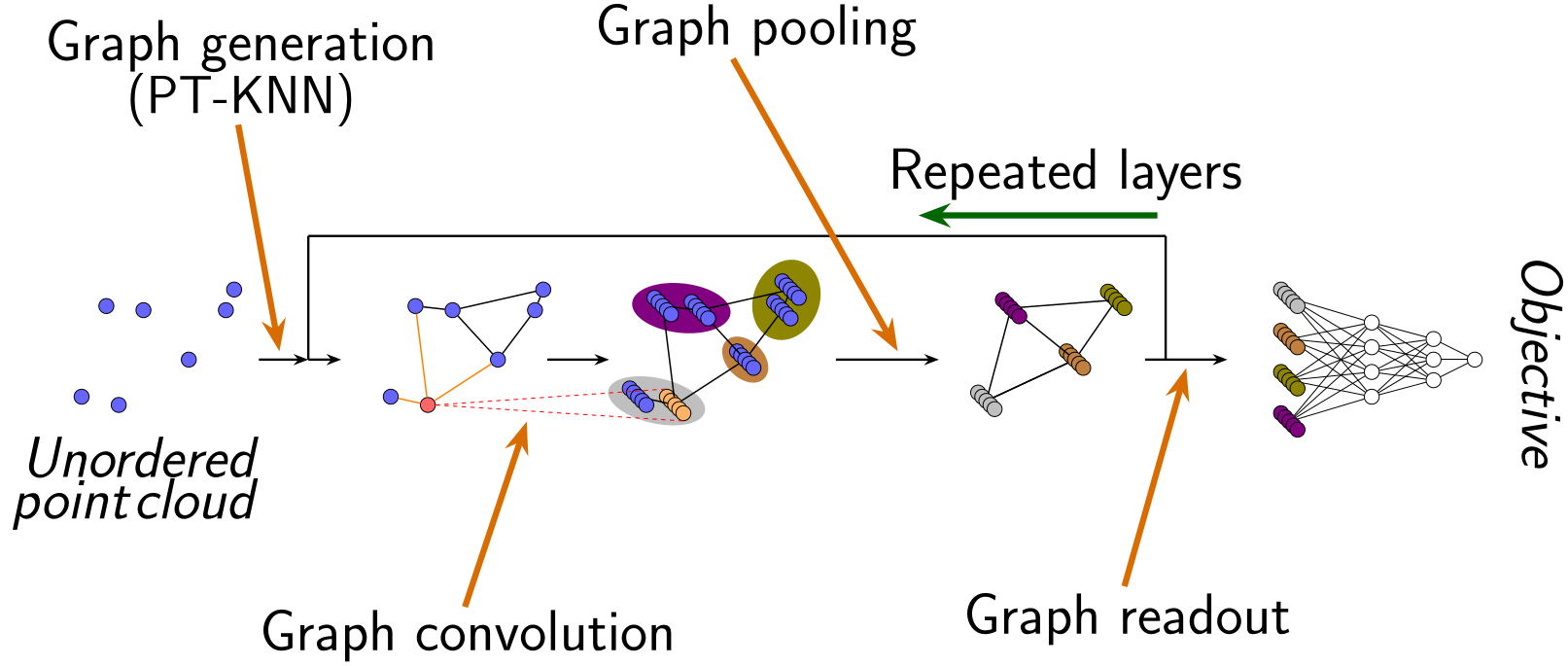}
				\caption{Full GNN pipeline. First is the graph construction from the point cloud data, using the PT-KNN algorithm. Then come several layers of convolution and pooling. A readout pooling layer flattens the graph, the output of which is fed to a final MLP.}\label{sec:full-fig:full_pipeline}
			\end{figure}

			The complete Graph Neural Network (GNN) pipeline designed for efficient particle identification is illustrated in Figure \ref{sec:full-fig:full_pipeline}.
			It is structured in sequential blocks, each fulfilling one of three objectives: graph construction, geometric feature extraction, and graph-level feature regression.
			
			The graph construction starts from point cloud data, extracted from the detector sensors, and computes edges through the KNN algorithm. If timing information is included/relevant, one might want to perform the KNN algorithm using space-time coordinates. It is however, algorithmic complexity wise, more efficient to use only space coordinates, and use the PT-KNN algorithm. This yields graphs with node features corresponding to the deposited energy in each cell, and potentially the cell position. The graph's edges represent connections between geometrical neighbourhoods, and carry as feature the edge length.
			
			Once a graph is obtained, it goes through repeated layers of convolution and pooling (CP layers). The convolution operation is given by equation \ref{eq:mp_generic_eq}. It collects the features from its neighbours through the message function $\phi$, and aggregates all messages to update its own features. Skip connections, which concatenate the outputs of previous layer to that of the current one, may be implemented to smoothen the learning gradients \cite{huang_17}. To do so, we simply concatenate the node's input features (layer $l=0$) to the output of convolution layer $l>0$. Convolution layers may be interleaved by pooling layers. They are implemented as the three steps of clustering, reduction and connection, using the $N$-Treclus algorithm for the clustering step. Note that the pooling layers make it impossible to implement skip connections with the input features, as clusters hold many of the input nodes. As such, if skip connections are implemented, we aggregate the input features of nodes, using the same aggregator as for the nodes' current features, so that each cluster has a single set of features in the input phase space.
			
			After repeated layers of convolution and pooling, we use the detector voxelisation to assign each node to its corresponding readout region. The features are aggregated within each region, and might concatenate skip connections from the CP layers to the readout tensor. If this is the case, the features from layer $l$ get carried through the subsequent pooling layers in the same way as previously described for the initial features. The flattened graph representation, with its fixed size, is then passed through an artificial neural network, that will map the geometric features extracted by the convolution to the regression output (e.g. particle ID, energy, etc.).
			
			This structure introduces many composing elements that are highly adapted for a parallelised implementation. If these algorithms are to be used in the resource and latency constraint heavy reconstruction environments of granular detectors like HGCAL or Hyper-Kamiokande, one must maximise concurrent processing.
			
			First, the graph construction using a $k$ nearest neighbours (whether it be with or without proximity tables) is inherently parallelisable thanks to the asymmetric assignment of neighbours. Many algorithms can then be implemented to ensure the resulting graph is undirected (i.e. $e_{vw}\in E\Leftrightarrow e_{wv}\in E$) \cite{bretto_12}. One might wish to use directed graphs, which is more efficient as it does not need symmetric neighbourhood assignment. This however hinders the free flow of information through the graph, needing bigger models and longer training.
			
			Similarly, graph convolution is highly parallelisable, as operations are decoupled across the graph. The calculation of localised messages along explicit edges and the subsequent aggregation depend strictly on node neighbourhoods. This allows for simultaneous node feature updates with optimised parallel kernels assigned to individual nodes.
			
			The parallelisation of Treclus is a trickier task, as the necessity of a graph matching make it inherently sequential. There has however been works for a threaded implementation of graph matchings, most notably for the Graclus clustering routine \cite{fagginger_12}, which makes matching proposals from each node on parallel threads, then checks for symmetric proposals, and repeat until no more proposals can be made.
			
			The readout pooling operation is also inherently parallel, the mapping from nodes to readout region providing a natural mapping to threads for parallel aggregation and writing to the readout tensor.
			
			Finally, from the locality of the graph convolution operation, one could consider an acceleration through the partitioning of the graph. The assumption is that paths between nodes are degenerate, and as such that partitioning the graph will have a limited impact on the information flow within individual partitions. Convolution and pooling layers may then be parallelly implemented on each individual partition of the graph, with potential reconnections between the CP layers to ensure the extraction of macro-scale features by the message passing convolution operation. We present a preliminary study of this approach in section \ref{sec:ergo_parallel}.
		
	\section{OPIF: Particle Identification Pipeline}\label{sec:opif}
	
		We first validate the algorithmic framework presented in the previous section by applying it to particle identification. We test it on a five-way particle identification problem, which will classify $e^-$, $\gamma$, $2\gamma$, $\pi^+$ and $\mu$ events. The OGCID Particle Identification Framework (OPIF) takes graphs built using the PT-KNN algorithm, and performs successive message passing convolution and pooling operations on it. The readout operation is then applied to flatten the graph to apply a final neural network operation to regress class probabilities, exploiting the specific densities of models.
		
		Furthermore, the design of OPIF serves as a functional demonstration of how these GNN pipelines can be implemented with varying degrees of specialisation, allowing the architecture to adapt its complexity based on the specific physical requirements of the reconstruction task.
		
		\subsection{Pipeline Structure and Training}\label{sec:opif-subsec:struct}
		
			The primary objective of the OPIF classifier is to accurately identify the incident particle responsible for a given calorimeter event, specifically distinguishing between electromagnetic showers ($e^-/\gamma$), photon pairs ($2\gamma$), hadronic showers ($\pi^+$), and minimally ionising particles ($\mu$).
			
			Graphs are obtained from raw events using the previously introduced Proximity Table KNN algorithm, using a proximity table containing all of the cells of the D2 detector and built from the 3D euclidean distance. We observed that the performance of the classifier would not improve by increasing $k$ above 4. As such, we compute the edges with the PT-KNN algorithm using  $k=4$. The resulting graphs have node features $V\in \mathbb{R}^+$, which correspond to the energy measured by each sensor, and their edge features are the euclidean distance between their end points. Each node also holds a set of hidden features, which corresponds to their position and readout region index. These features are not used in the convolution operation, but will be used throughout the pipeline, for pooling and readout. The graphs are then passed through successive layers of graph convolution and pooling layers. The convolution operation is the one presented in equation (\ref{eq:mp_ours}). As this is a classification task, the aggregator $\square$ computes the feature-wise maximum of all extracted messages \cite{goodfellow_16}, to extract locally protruding features. The pooling implements the $N$-Treclus algorithm, using the 30-th percentile of edge lengths for the threshold value, so as to have roughly 10 nodes per populated readout region, ensuring information travelled through the entire graph. Once again, the reduction operation computes the feature-wise maximum of nodes that are within the same cluster. A random node within the cluster is taken to be the resulting node's position. The hidden features (position and readout region index) for this new node correspond to the randomly selected node within each cluster. We then use the node positions of the pooled graph, and re-compute the edge lengths for the subsequent convolution operation.
			
			After the convolution and pooling layers, we perform a readout operation on the overall graph by dividing each subdetector (ECAL, near HCAL, far HCAL) and aggregating the features of nodes within the same region together, using the maximum operator once again. The readout region index for each cell is computed beforehand to avoid the overhead induced by computing the readout region at each forward pass and during inference. The subdivision in readout regions of the detector was designed to respect the detector's angular geometry, as described in section \ref{sec:readout}.
			
			The tensor resulting from the readout operation is then given as input to an MLP, the last layer of which is a softmax operation to normalise the probabilities vector. 
			
			Another classifier is added for showers classified as electromagnetically decaying events, to distinguish $e^-$ and $\gamma$ events. It follows exactly the same structure as the previous model, and allows to have a greater classification power, giving a five-way classifier which may distinguish electrons, $\gamma$s, $\pi^0\to 2\gamma$, $\pi^+$s and muons. 
			\begin{figure}
				\centering
				\includegraphics[width=0.75\textwidth]{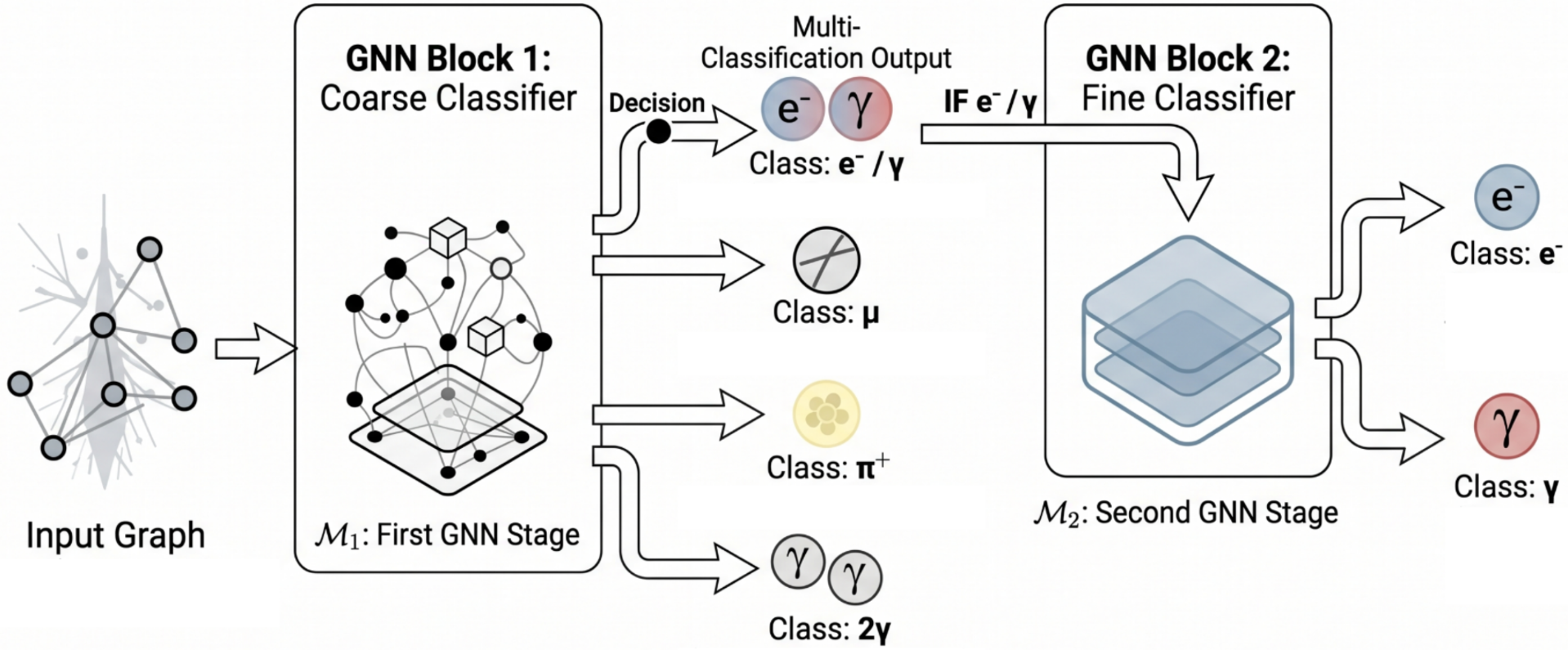}
				\caption{The OPIF classification pipeline. It is a two step classifier, first dealing with easy to distinguish EM, MIP, hadronic and 2$\gamma$ classes, and classifying $e^-$s and $\gamma$s in a second, finer classifier.}
			\end{figure}
			The choice of the two-step classification was justified by the difficulties a general classifier had to distinguish electrons and photons. In particular, we observed that the model would often collapse to a random classifier for events in the ECAL. This is explained by the fact that the scale at which the coarser classifier works (full detector size) is much greater than that at which a photon and an electron shower are distinguishable, at the very beginning of the shower. Furthermore, this allows for much smaller models, as they do not need to learn at both scales, greatly improving training and inference times. This second block also allows us to add a finer division of the ECAL for the readout operation, needed to distinguish EM showers. Indeed, the most notable difference between $e^-$ and $\gamma$ showers is the shower start, which leaves no track before the shower triggers for the photons.
			
			The first GNN block is a model with $\sim 350,000$ parameters, with 3 convolution layers, each using a 3 layer MLP for the message function. It was trained on a set of 2M showers, equally divided into each of the output classes (500,000 graphs each). The joint $e^-/\gamma$ class is evenly split between both of these incident particle (250,000 graphs each). The model being quite fast to converge, we train this model for 100 epochs using the Cross Entropy Loss, with a learning rate of $10^{-4}$ and an Adam optimiser \cite{kingma_14}. 
			
			The second GNN block, which classifies photons and electrons, has around 1.2 M parameters, with 6 convolution layers, each using a 3 layer MLP for the message function. It was trained on 2M events from each particle class $e^-$ and $\gamma$. The model trained for 500 epochs with a Binary Cross Entropy Loss and Adam optimiser, setting the learning rate to $10^{-4}$.

		\subsection{Classifier Performance}
		
			The performance of the OPIF classifier is evaluated using a global confusion matrix, and the Receiver Operating Characteristic (ROC) curves, varying the acceptance threshold to see the evolution of the signal purity. These metrics allow for a quantitative assessment of the model's ability to distinguish between the five particle classes across the different energy ranges (10 to 100 GeV) and incident angles ($0^\circ$ to $15^\circ$) provided in the D2 dataset.
			
			The first stage of the pipeline, GNN Block 1, handles the four-way classification between electromagnetic showers ($e^-/\gamma$), muons ($\mu$), charged pions ($\pi^+$), and photon pairs ($2\gamma$). 
			
\begin{figure}
				\centering
				\begin{subfigure}{0.48\textwidth}
					\centering
					\includegraphics[width=\textwidth]{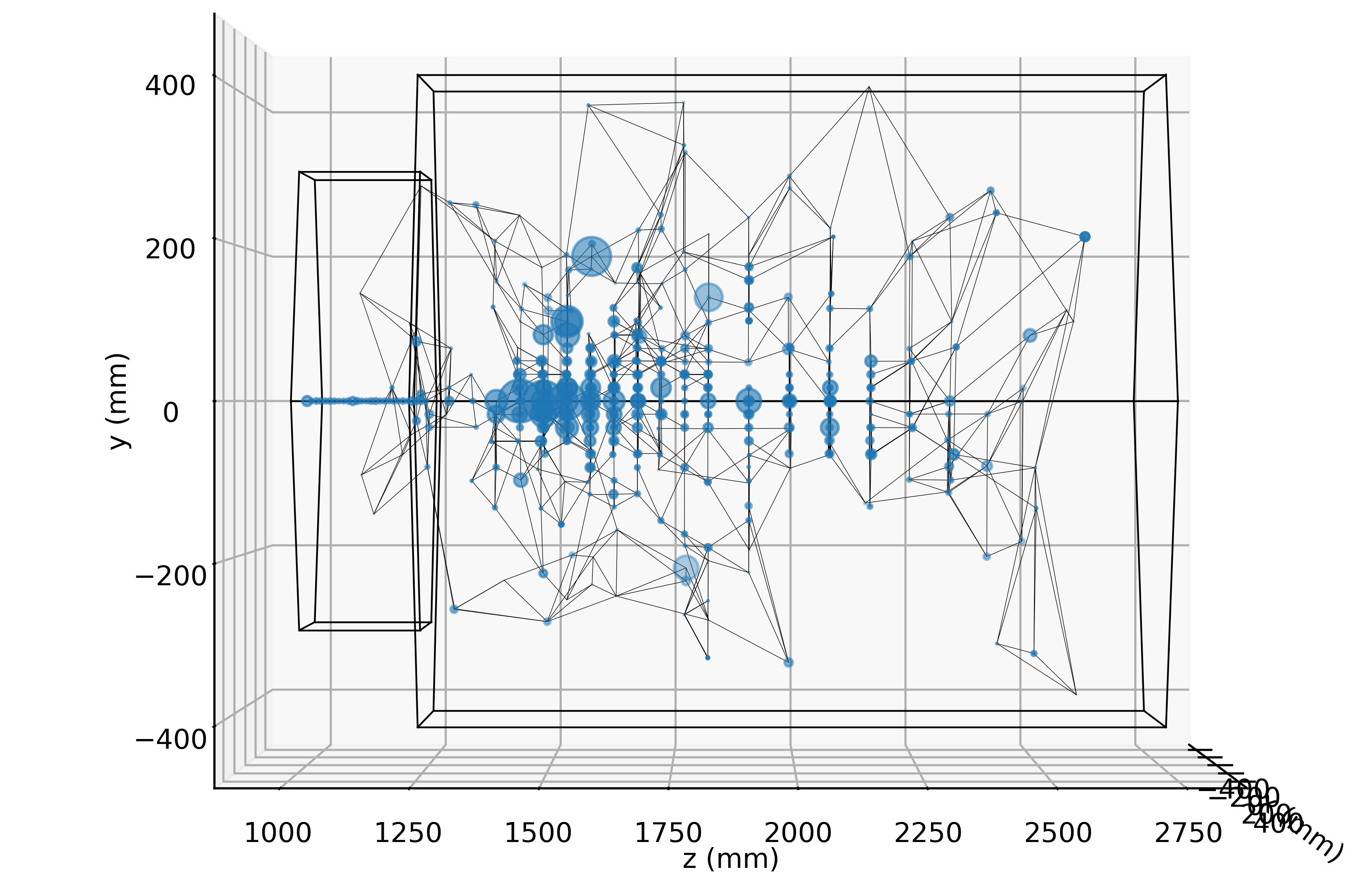}
					\caption{Late showering $e^-$, triggering a hadronic shower.}\label{sec:opif-subsec:perf-fig:outliers-subfig:hadronic_elm}
				\end{subfigure}
				\hfill
				\begin{subfigure}{0.48\textwidth}
					\centering
					\includegraphics[width=\textwidth]{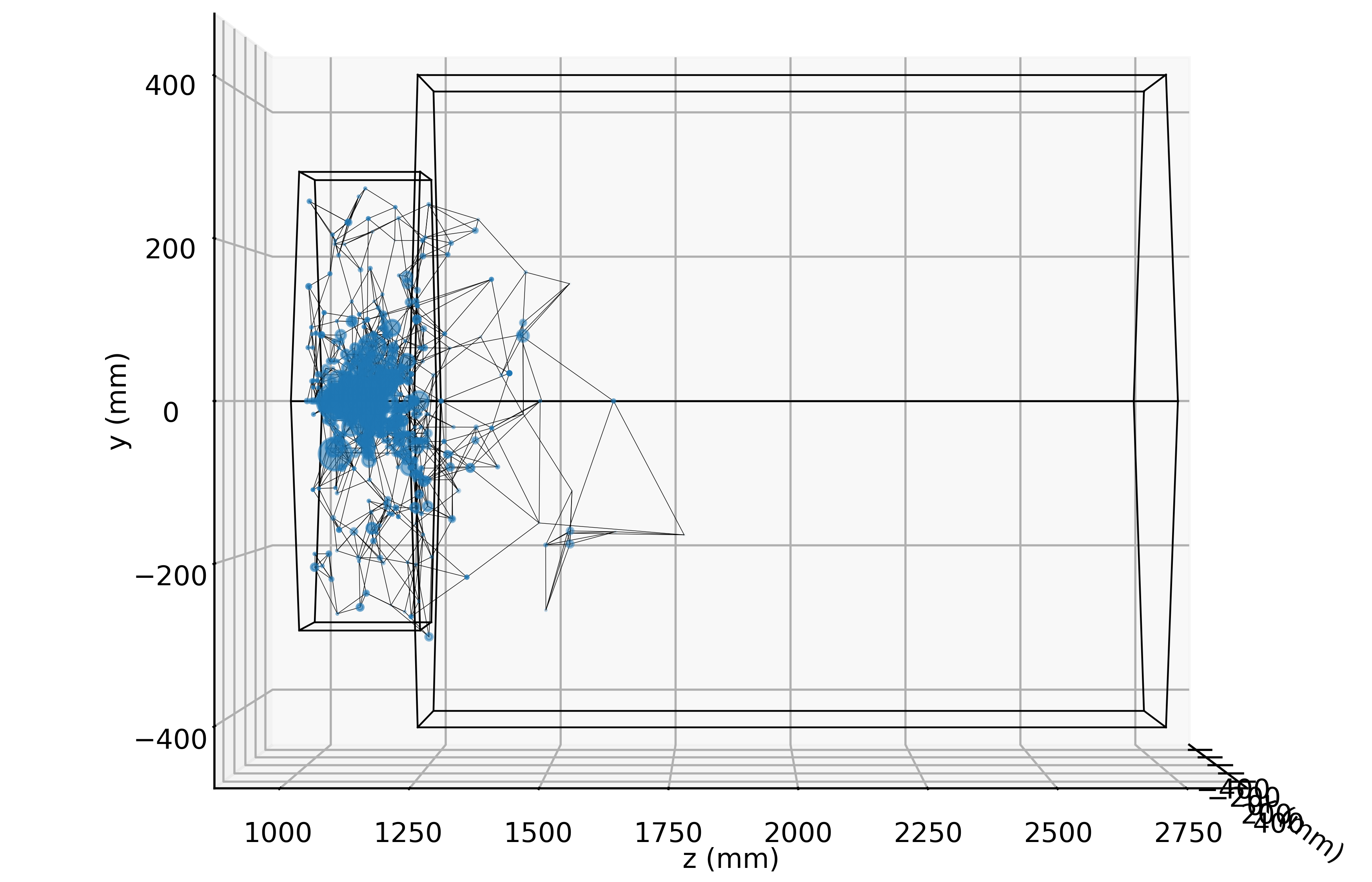}
					\caption{$\pi^+$ interacting with the ECAL, triggering a set of $\pi^0\to 2\gamma$ showers.}\label{sec:opif-subsec:perf-fig:outliers-subfig:em_pip}
				\end{subfigure}
				\caption{Outlying events in the D2 dataset for the particle identification task.}\label{sec:opif-subsec:perf-fig:outliers}
			\end{figure}

			Part of this task is relatively simple: if we look at the event displays from Figure \ref{sec:opif-subsec:perf-fig:outliers}, we understand that the discrimination between $\{e^-/\gamma, 2\gamma\}$, muons and hadrons is a basic density task, mostly dealt with by the readout operation. The convolution layers are however still useful, in particular for the outlying events. Indeed, we observed that applying only the readout operation performed a little bit worse than with the convolution layer. This is in particular true for particles that shower with atypical profiles, such as $\pi^+$ that interact with the ECAL and produce $\pi^0$s, themselves decaying in the ECAL (Figure \ref{sec:opif-subsec:perf-fig:outliers-subfig:em_pip}).
			In such a case, the understanding of the geometry of the early shower is essential to a successful classification. It is also necessary to have an understanding of the geometric structure of the shower for the $e^-/\gamma$ versus $2\gamma$ classification. Indeed, formulating the task as a density problem, these two classes are virtually identical. The convolution layer allows the model to make this distinction.
			
\begin{figure}
				\centering
				\includegraphics[width=0.5\textwidth]{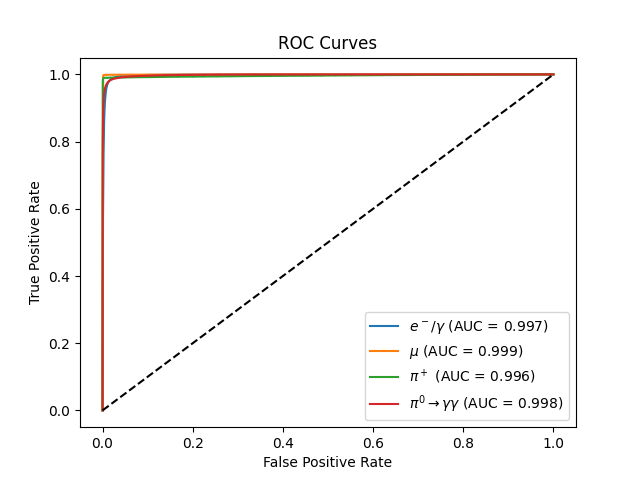}
				\caption{Receiver operating characteristic (ROC) curve for the first classifying GNN of the OPIF pipeline. We observe a close to perfect performance for muon and pion classes, while the EM classes ($e^-/\gamma, 2\gamma$) have a less polarised output.}\label{sec:opif-subsec:perf-fig:roc1}
			\end{figure}

			We first look at the ROC curve for the 4-class PID task performed by the first, coarse GNN. The ROC for each class are shown in Figure \ref{sec:opif-subsec:perf-fig:roc1}.
			We observe that for $\mu$ and $\pi^+$ classifier, the model is more sure of itself, with less progressive score values. For the ECAL events however, the model returns more progressive scores, due to the greater proximity between these event classes, making the task trickier. This is clearer when looking at the confusion matrix (Table \ref{sec:opif-subsec:perf-tab:confusion1}), obtained by tagging events based on the highest prediction probability.
			\begin{table}
				\centering
				\begin{tabular}{cc||cccc}
					\multicolumn{2}{c||}{\multirow{2}{*}{}}         & \multicolumn{4}{c}{Predicted} \\ \cline{3-6} 
					\multicolumn{2}{c||}{}                          & $e^-/\gamma$     & $\mu$     & $\pi^+$     & $2\gamma$     \\ \hline\hline
					\multicolumn{1}{c|}{\multirow{4}{*}{True}} & $e^-/\gamma$ & 97.91 \%     & $<$0.01 \%    & 0.10 \%     & 1.99 \%     \\
					\multicolumn{1}{c|}{}                      & $\mu$ & 0.01 \%     & 99.75 \%     & 0.24 \%     & $<$0.01 \%     \\
					\multicolumn{1}{c|}{}                      & $\pi^+$ & 0.35 \%     & 0.76 \%     & 98.87 \%     & 0.02 \%     \\
					\multicolumn{1}{c|}{}                      & $2\gamma$ & 3.01 \%     & $<$0.01 \%     & 0.03 \%    & 96.95 \%    
				\end{tabular}
				\caption{Confusion matrix for the first layer of the classification model. We observe that the biggest error comes from misclassifying particles showering in the ECAL.}\label{sec:opif-subsec:perf-tab:confusion1}
			\end{table}
			
			To evaluate the classification power of the second GNN block in isolating single electrons from photons, we examine its dedicated ROC curve and confusion matrix. As with the coarse classifier, this performance is assessed across the full spectrum of simulated energies (10 to 100 GeV) and incident angles ($0^\circ$ to $15^\circ$).
			\begin{figure}
				\centering
				\includegraphics[width=0.5\textwidth]{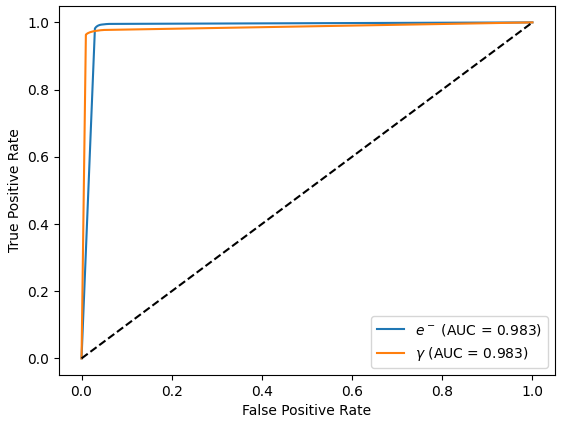}
				\caption{ROC curve for the second classifier ($e^-; \gamma$). One observes that due to the bigger difficulty of the task, in comparison with the first layer, the AUCs are lower than for the 4-class PID model. They however remain more than satisfactory.}\label{sec:opif-subsec:perf-fig:roc2}
			\end{figure}
			We observe that the ROC curves are not as good as for the previous layer of the model (see Figure \ref{sec:opif-subsec:perf-fig:roc1}). This is because the task is much more difficult. We however achieve a good classifying performance, as shown by the confusion matrix in Table \ref{sec:opif-subsec:perf-fig:roc2}, obtained by taking the most probable class, as predicted by the model ($p_\text{PID} > 0.5$).
			\begin{table}
				\centering
				\begin{tabular}{cc||cc}
					\multicolumn{2}{c||}{\multirow{2}{*}{}}                   & \multicolumn{2}{c}{Predicted} \\ \cline{3-4} 
					\multicolumn{2}{c||}{}                                    & $e^-$   & $\gamma$       \\ \hline\hline
					\multicolumn{1}{c|}{\multirow{2}{*}{True}} & $e^-$ & 98.75 \%      & 1.25 \%      \\
					\multicolumn{1}{c|}{}                      & $\gamma$ & 3.33 \%       & 96.67 \%
				\end{tabular}
				\caption{Confusion matrix for the EM single shower event classifier.}\label{sec:opif-subsec:perf-tab:confusion2}
			\end{table}
			The slightly decreased signal purity, when compared to the previous layer, is explained by the similarities in shower development for photons and electrons. Indeed, the main difference appear at the beginning of the shower, and in particular by the mean free path of the particle before it triggers a shower. The variability of the path of photons before pair production also explained the decreased signal purity of the photon class.
			
			Putting both pipelines together, we obtain a full classifier for the D2 dataset, with $e^-, \gamma, \mu, \pi^+$ and $2\gamma$ events. The resulting confusion matrix is presented in Table \ref{sec:opif-subsec:perf-tab:confusion12}.
			\begin{table}
				\centering
				\begin{tabular}{cc||ccccc}
					\multicolumn{2}{c||}{\multirow{2}{*}{}}                 & \multicolumn{5}{c}{Predicted}                                                     \\ \cline{3-7} 
					\multicolumn{2}{c||}{}                                  & $e^-$              & $\gamma$ & $\mu$              & $\pi^+$  & $2\gamma$         \\ \hline \hline
					\multicolumn{1}{c|}{\multirow{5}{*}{True}} & $e^-$     & 96.69 \%           & 1.22 \% & $<$0.01 \% & 0.12 \%  & 1.97 \%           \\
					\multicolumn{1}{c|}{}                      & $\gamma$  & 3.26 \%           & 94.65 \% & $<$0.01 \% & 0.10 \%  & 1.99 \%           \\
					\multicolumn{1}{c|}{}                      & $\mu$     & $<$0.01 \% & $<$0.01 \%  & 99.75 \%           & 0.24 \%  & $<$0.01\% \\
					\multicolumn{1}{c|}{}                      & $\pi^+$   & 0.20 \%            & 0.15 \%  & 0.76 \%            & 98.87 \% & 0.02 \%           \\
					\multicolumn{1}{c|}{}                      & $2\gamma$ & 1.54 \%            & 1.48 \%  & $<$0.01 \% & 0.03 \%  & 96.95 \%         
				\end{tabular}
				\caption{Confusion matrix for the full OPIF classification pipeline. We observe a good signal purity, with the largest errors coming from the EM-shower classification.}\label{sec:opif-subsec:perf-tab:confusion12}
			\end{table}
			
			We observe a decreased classification accuracy for $e^-$ and $\gamma$ events. This is due to the propagation of errors from the first GNN model to the second one. This is however a satisfactory performance for this difficult task.
			
			A comparison to the classification performances of TICL \cite{cms_22_track_classif, pilato_20}, despite the very different contexts, confirms that our approach is very promising for the particle identification task. The performance could be improved by the incorporation of these pipeline blocks within other reconstruction models, most notably TICL. In particular, one could use the previously presented pipelines using graphs built from partially reconstructed clusters obtained by TICL, which we suspect to improve performance, while also reducing the size of the model.
	
	\section{ERGO: Particle Energy Regression Pipeline}\label{sec:ergo}
		
		We then test our framework for the regression of the energy of the incoming particles. We introduce the ERGO (Energy Regression on Graph Objects) pipeline. It exploits a similar structure to the one used for the coarse OPIF GNN pipeline. We also use ERGO to study the parallelisation possibilities of our pipeline structure.
		
		\subsection{Model Structure}\label{sec:ergo_struct}
			
			The ERGO model takes particle showers from the D2 dataset, and regresses the energy of the incident particle from the shower. Since the regressor looks at very different particle shower signatures, we train multiple energy regression models, one for each particle type, to learn the specific properties of each showering mechanism. Note that because muons traverse the detector as minimum ionising particles (MIPs), their energy is virtually irresolvable solely using the D2 detector, which does not have an external magnetic field. The $\gamma\gamma$ events are also problematic, as we are interested in obtaining the energy of the individual photon showers, and further treatment needs to be done before we can regress their respective energies (see section \ref{sec:soop}).
			
			As a result, for now, we only look at two models, one to jointly regress $e^-$ and $\gamma$ energies, the other to regress the $\pi^+$ energies. They  are both based on the same model structure, adapted to each problem.
			
\begin{figure}
				\centering
				\includegraphics[width=0.5\textwidth]{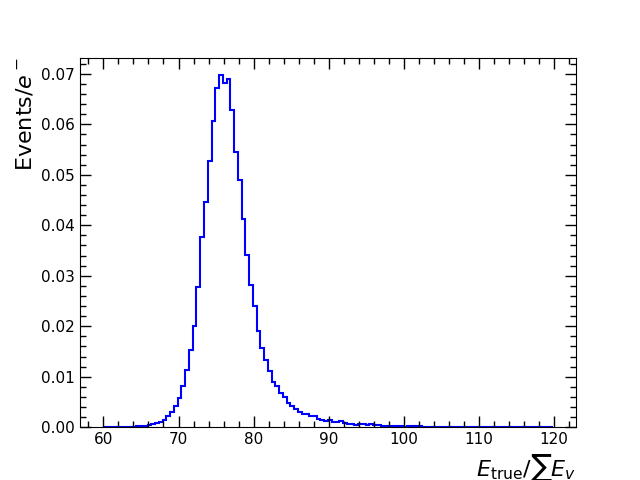}
				\caption{Normalisation of the regression objective to $E_\text{true}/\sum_vE_v$. This changes the energy distribution from a uniform to a normal distribution, which are easier for deep learning models to learn.}\label{sec:ergo-fig:energy_normalisation}
			\end{figure}

			This model basis is very similar to that used for the OPIF classifier: Graphs are generated using the PT-KNN algorithm, using $k=4$, for the same reason as for the OPIF classifier. The result are graphs with node features $v\in V\subseteq R^+$, which correspond to the energy $E_v$ that was deposited in the corresponding cell. Each hit also holds its position and readout region index as a hidden feature, not used by the convolution operations, but used for the readout pooling operation. The edge features $e_{vw}\in E\subseteq \mathbb{R}^+$ correspond to the distance between nodes $v, w \in V$. The graph then goes through several layers of convolution and pooling. Since this is a regression task, we use a mean operator as the $\square$ aggregator \cite{goodfellow_16}. Convolution layers alternate with pooling layers, which are implemented using the $N$-Treclus algorithm, using as threshold a variable quantile of edge lengths. Within each cluster, we choose a node at random, and combine all nodes from that cluster to this node position, ensuring the embedding of nodes on a predefined grid, which is later used for the readout. The features of the resulting node are the featurewise sum of all nodes within the cluster. The sum operation is chosen to keep the information carried by the number of nodes inside a cluster. Indeed, if we used the average of the features, the output would have the same order of magnitudes for clusters containing 2 nodes, and those containing 20 nodes. This is of course not desirable, as hit density carries information on the shower. After the reconnection step, during which the new clustered nodes inherit the adjacencies of their constituting nodes, the edge lengths are re-calculated. After the repeated convolution and pooling layers, we apply a readout operation. It is the same as the one used for the OPIF models, which respects the detector symmetries: the detector is sliced longitudinally and divided in concentric regions (fig. \ref{sec:readout-fig:readout_regions}), and the nodes within each voxel are pooled using a featurewise sum pooling. Because of all the sums of features performed during the above layers of the model, the resulting features might have become excessively big, which is problematic for the gradient descent algorithm \cite{bjorck_18}. To solve this, a batch norm layer is implemented to the readout tensor, before it is fed into an artificial neural network. The output corresponds to the ratio given as
			\begin{equation}
				\frac{E_\text{pred}}{\sum_v E_v},
			\end{equation}
			with $E_\text{pred}$ the predicted energy. This allows a normalisation of the uniform distribution of true energies $E_\text{true}$, as shown by Figure \ref{sec:ergo-fig:energy_normalisation}.
			This is done because deep networks are best at regressing variations from an average behaviour (i.e. normal distribution) than a uniform output \cite{white_89}.
					
			The pipelines for energy regression of $e^-/\gamma$s and positive pions both use the above structure. We however use a slightly deeper model for the pion energy regression, by implementing deeper message functions. This is done to better learn the more complex dynamics of hadronic showers. Indeed, while all of the EM shower energy is deposited in the detector, ensuring that $E_\text{true} \sim A \sum_vE_v$, some of the hadronic shower elements are not visible by the calorimeter, taking away energy that will not be measured. We also adapt the readout to the task: an electron energy regression does not need to have a very granular readout in the HCAL, which it will never reach. Similarly, $\pi^+$s will mostly traverse the ECAL without showering. This allows us to limit the MLP's size while still increasing the readout region density in relevant areas.
			
		\subsection{Training}
			
			The $e^-/\gamma$ energy regression model was trained using $10^6$ graphs, evenly split between the two types of particles. The model had roughly $10^5$ parameters, with 4 convolution and pooling layers. The messages for the convolutions were MLPs with 3 hidden layers. It was trained using a Mean Squared Error Loss (MSE Loss), with a learning rate of $10^{-3}$ and an Adam optimiser. The model converged after $\sim 500$ epochs.
			
			The pion energy regressor had $\sim 5\times 10^5$ trainable parameters, with 7 convolution and pooling layers, each  with messages with 3 hidden layers . That model was trained on $1.5\times 10^6$ graphs. Once again, we use the MSE loss, with a learning rate of $10^{-4}$ for the Adam optimiser.
		
		\subsection{Performance}
		
			The model's response is evaluated as the ratio
			\begin{equation}
				R_\text{ERGO}=\frac{E_\text{pred}}{E_\text{true}},
			\end{equation}
			providing a normalised metric across all energy ranges. If we look at the distribution of this response for both regression models, we observe a Gaussian distribution with a low standard deviation, of at most 3 GeV for the electrons and 7 GeV for pions.
			\begin{figure}
				\centering
				\begin{subfigure}{0.49\textwidth}
					\centering
					\includegraphics[width=\textwidth]{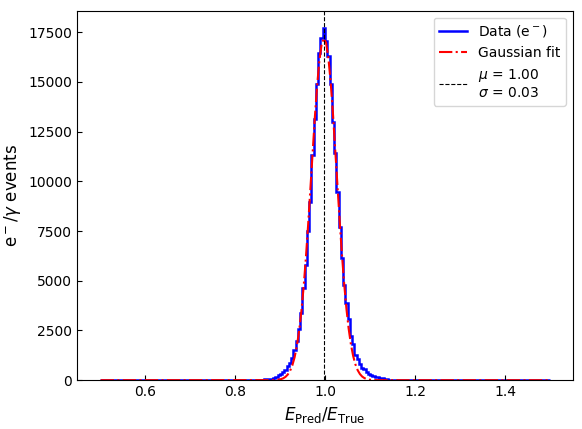}
				\end{subfigure}
				\begin{subfigure}{0.49\textwidth}
					\centering
					\includegraphics[width=\textwidth]{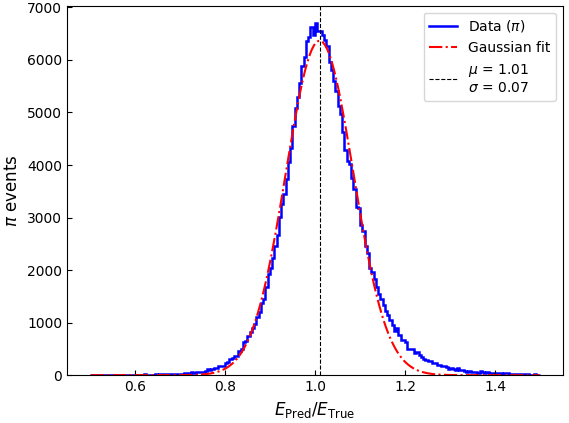}
				\end{subfigure}
				\captionof{figure}{Energy response $E_\text{pred} / E_\text{true}$ for $e^-/\gamma$ (left) and $\pi^+$ (right) showers. We observe Gaussian distribution, as expected, with a low standard deviation of less than 7 GeV.}
			\end{figure}
			While this is a good estimation of the model's performance, most relevant is the energy resolution $\sigma / \mu$. Binning in energy, we look at the distribution of outputs. We fit the distribution of responses with the Cruijff function, defined as
			\begin{equation*}
				f(x; A, \mu, \sigma_{L,R}, \alpha_{L,R}) = A \exp \left( -\frac{\left(x-\mu\right)^2}{2\sigma^2 + \alpha\left(x-\mu\right)^2} \right),
			\end{equation*}
			where $\sigma$ defines the left and right standard deviation, and $\alpha$ the left and right distribution tails. They are defined by
			\begin{equation*}
				\sigma = \begin{cases}
					\sigma_L &\text{if }x<\mu, \\
					\sigma_R &\text{if }x\geq\mu,
				\end{cases} \qquad \text{and} \qquad \alpha = \begin{cases}
					\alpha_L &\text{if }x<\mu, \\
					\alpha_R &\text{if }x\geq\mu.
				\end{cases}
			\end{equation*}
			
\begin{figure}
				\centering
				\begin{subfigure}{0.48\textwidth}
					\centering
					\includegraphics[width=\textwidth]{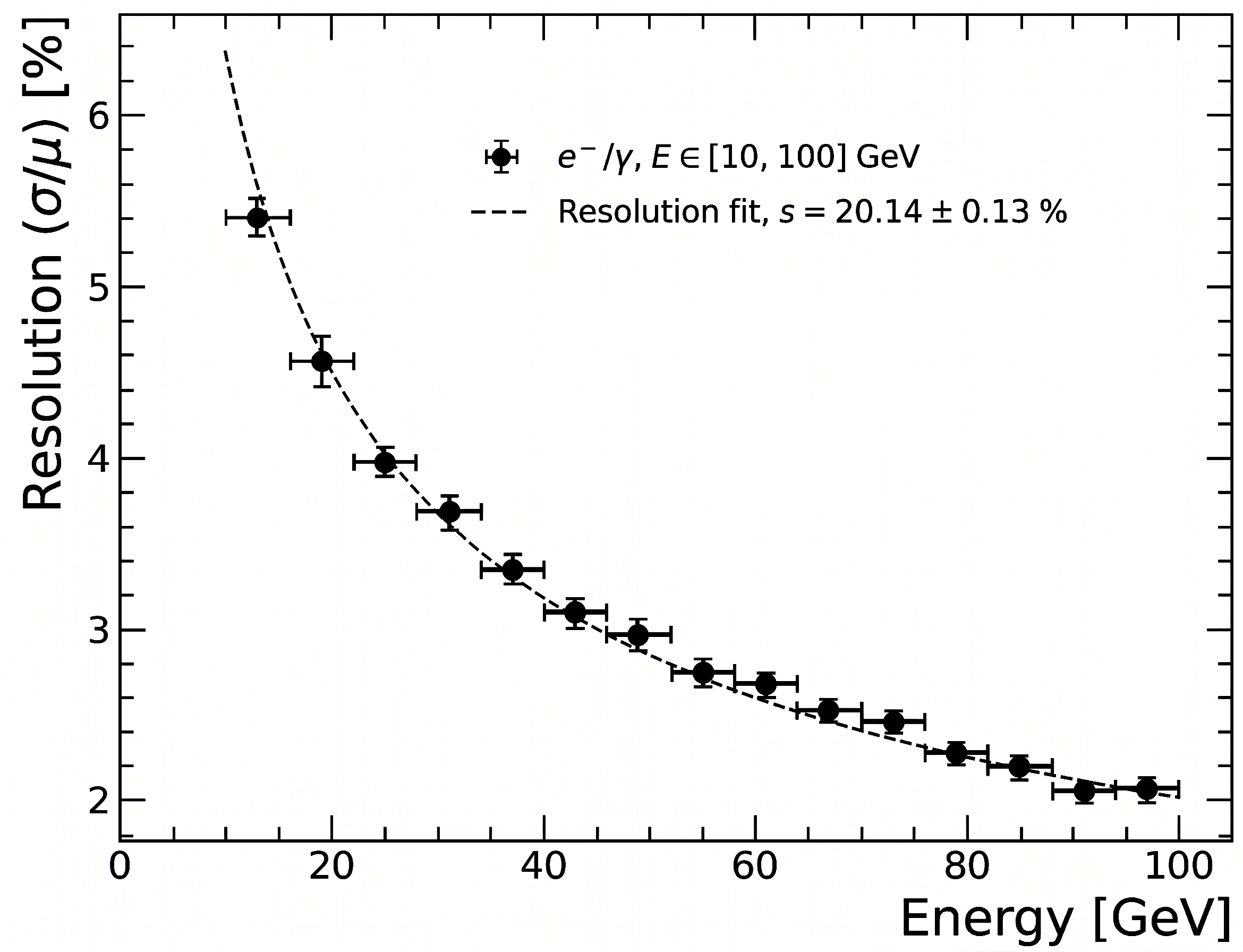}
					\caption{$e^-/\gamma$ energy resolution.}
				\end{subfigure}
				\begin{subfigure}{0.49\textwidth}
					\centering
					\includegraphics[width=\textwidth]{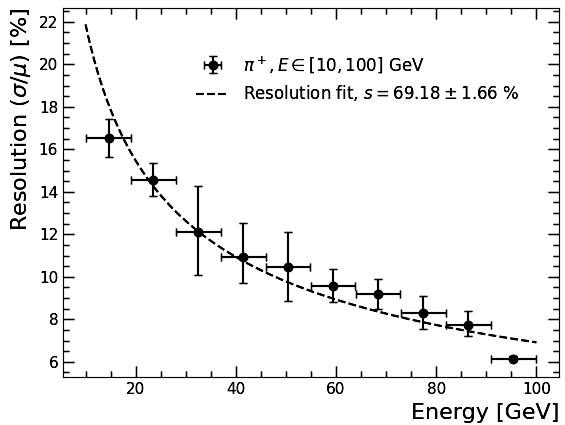}
					\caption{$\pi^+$ energy resolution.}
				\end{subfigure}
				\caption{Energy resolutions for the $e^-/\gamma$ and $\pi^+$ energy regressors. Fitting with equation (\ref{sec:ergo-eq:res}) with null noise and constant terms, we obtain $S_\text{EM}=20.14\pm 0.13$ \%, which are close to the theoretical value for the D2 ECAL, $S_\text{ECAL} \sim 20$ \%,  and $S_\text{Hadron} = 69.18\pm 1.66$ \%, which also is in the expected energy range $[50,80]$ \%.}\label{sec:ergo-fig:egamma_res}
			\end{figure}

			Taking the average standard deviation $\sigma = \frac{\sigma_L + \sigma_R}{2}$ for the resolution, we expect it to display the following dependence on the incident particle's energy \cite{fabjan_03}:
			\begin{equation}\label{sec:ergo-eq:res}
				\frac{\sigma}{E} = \frac{S}{\sqrt{E}} \oplus \frac{N}{E} \oplus C,
			\end{equation}
			where $\oplus$ is the quadratic sum. $S$ is the stochastic term, describing the physical fluctuations in shower development, $N$ the noise term, describing the electronic noise from the readout chain and $C$ the constant term, describing systematic noise. In the simulation for the D2 dataset, there is no noise emulation. As such, we consider $N=C=0$. The electron resolution is shown in Figure \ref{sec:ergo-fig:egamma_res}.
			We observe that the obtained stochastic term is $S=20.14\pm 0.13$ \%, which is in the expected theoretical range $S_\text{ECAL}\sim 20$ \%. For pions, we recover a stochastic term of $S_\text{HCAL}=69.18\pm 1.66$ \%, which is within the expected range $[50, 80]$ \%. This performance compares to the performance obtained on a beam test HGCAL prototype, when using a GNN model that dynamically recomputes the edges at inference \cite{cms_22_pip_res}, which is very computationally costly. While the two detectors and contexts are not comparable as is, the fact that we recover similar performances using our optimised model fully validates our approach.
		
		\subsection{Parallelising the Pipeline}\label{sec:ergo_parallel}
		
			We implement a parallelised version of the ERGO model, with the objective of reducing the inference time, in line with the goal of implementing GNN reconstruction elements for online triggering in the detector.
			
			The basic principle behind our parallelisation approach is that GNNs extract data from the graph by progressively going from a very local scale (node-level) to a general scale (graph-level). We try to exploit this progression, and build a two stage model. First, it extracts data locally, which is then collected and exploited at the graph level. This allows in particular to have the most part of the work performed by local convolution operations in a parallelised fashion, and then using a relatively small model for feature extraction at a larger scale. It also reduces the average number of operations for graph construction, as it reduces the size of the proximity table and list of potential neighbours.
			
\begin{figure}
				\centering
				\includegraphics[width=0.65\textwidth]{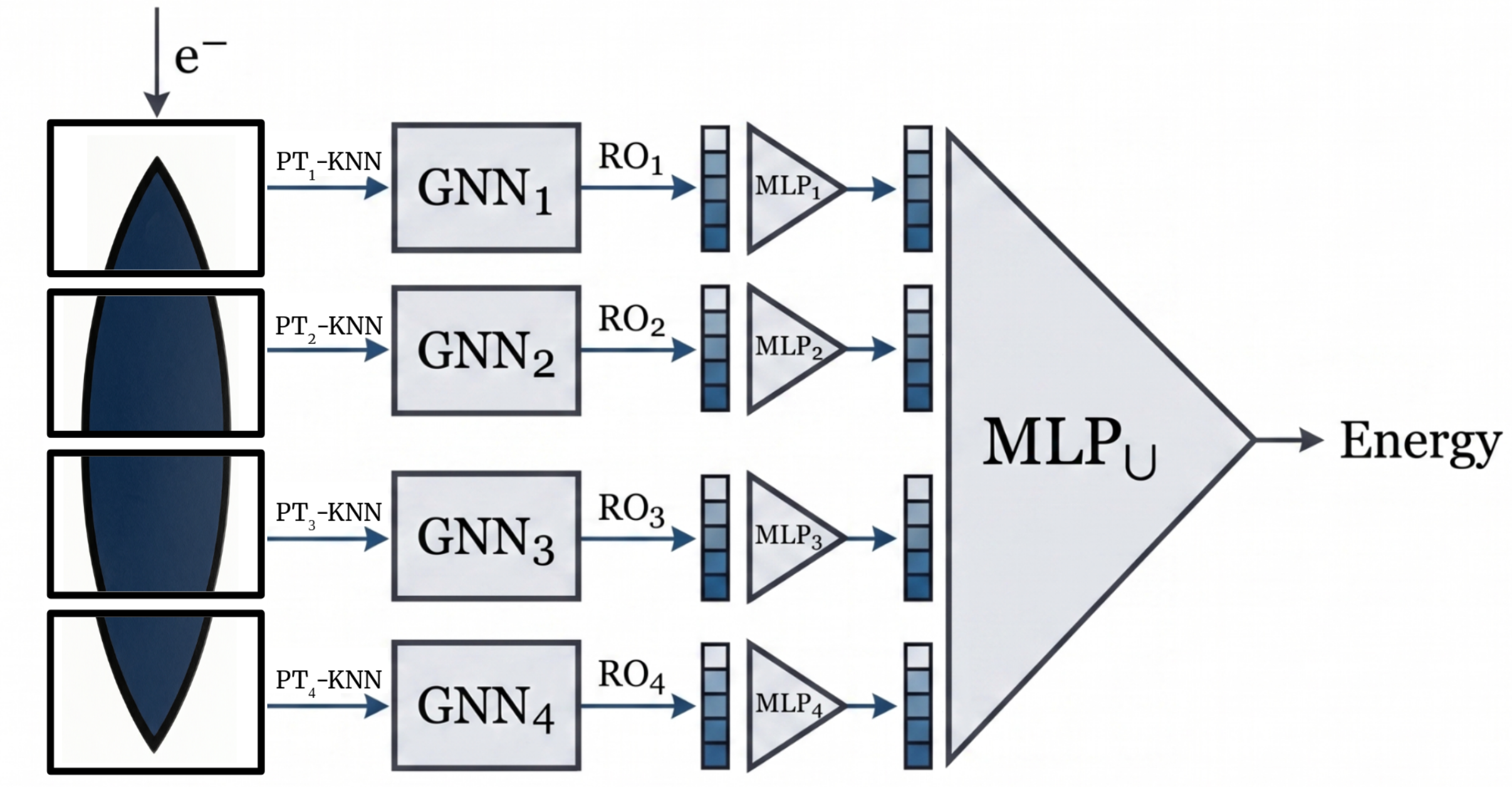}
				\caption{Parallelisation approach for the ERGO pipeline. Features are extracted from independently for each subregion of the detector, and recombined together by the final MLP.}\label{sec:ergo-subsec:parallel-fig:pipeline}
			\end{figure}

			The most basic way of doing this is by dividing the detector in subsections, and to consider each section independently from one another. A recombination model is then applied to the features extracted from each subregion to obtain a graph-wide result, as illustrated in Figure \ref{sec:ergo-subsec:parallel-fig:pipeline}.
			Within each subregion of the detector, we build a graph from the raw hits. This graph is then passed through layers of convolution and pooling, which follow the same structure as the pipeline previously presented in section \ref{sec:ergo_struct}. We then preform a readout, within each region, that follows the same division of the detector as with OPIF and ERGO (fig. \ref{sec:readout-fig:readout_regions}). The output of the readout layers are then passed to MLPs, one for each subregion.
			Note that the trainable parameters for each subsection of the detector are not shared amongst the pipelines, allowing each to specialise in the partial reconstruction of features for its specific part of the detector.	All of these operations can of course be ran parallelly for each subregion. The reduced number of nodes within each graph also lowers the number of operations to be performed at each convolution step. Once the local features for each subsection were extracted, we recombine them by inputting them into an MLP for the whole detector. The output of this MLP was, here, chosen to be the energy of the incident particle, as a metric for performance, but could be a set of features used to decide if the event is worth triggering on.
			
			In our implementation, we consider this problem for single electron events. The detector subregions are defined as longitudinal slices of the detector, i.e. each subregion is defined by a $z_\text{min}$ and $z_\text{max}$ value. Since most of the hits are within the ECAL of the detector, we will only define subregions for the ECAL, and consider the HCAL as one, big, subregion. It is kept as there might be hits inside it, and we wish not to loose any information with respect to the baseline, which uses the full detector.
			
\begin{figure}[tb]
				\centering
				\includegraphics[width=0.5\textwidth]{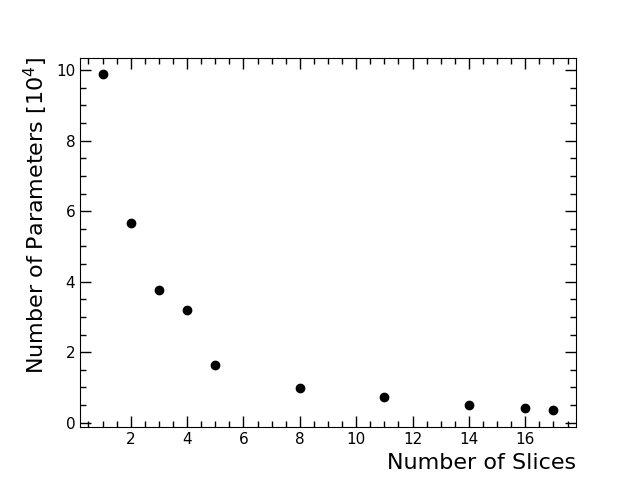}
				\caption{Number of parameters per slice ERGO sub-model as a function of the number of slices. We observe an exponential decrease in the number of parameters. This is a useful property as it means slicing the detector drastically reduces the number of parameters per model.}\label{sec:ergo_parallel-fig:param_count}
			\end{figure}

			The models for each slicing are defined so that the number of parameters in the full model remain in the order of $\sim 10^5$ parameters. As a result, the number of parameters per slice sub model is drastically reduced. Figure \ref{sec:ergo_parallel-fig:param_count} shows the number of parameters per GNN pipeline (specific to a single slice) as a function of the number of slices.
			
			We trained the model on $10^6$  electron and photon events. The loss curve is once again the Mean Squared Error Loss, using an ADAM optimiser with a $10^{-3}$ learning  rate for all slice counts. All models converged within 500 epochs.
			
\begin{figure}[tb]
				\centering
				\includegraphics[width=0.5\textwidth]{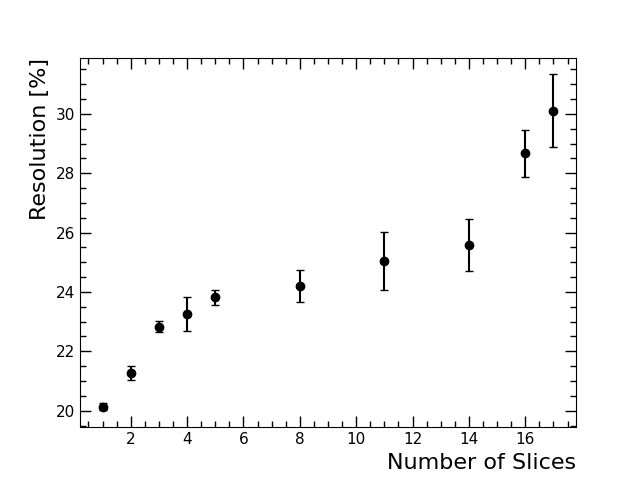}
				\caption{Evolution of the GNN's energy resolution dependent on the number of distinct slices in the ECAL.}\label{sec:ergo_parallel-fig:res_v_slice}
			\end{figure}

			As we did  with the full detector, we  look at the energy resolution, and use this as a metric for performance degradation as we increase the number of slices in the ECAL. This evolution is presented in Figure \ref{sec:ergo_parallel-fig:res_v_slice}.
			We observe a quick decrease in the resolution of the model. This is likely due to the loss of communication between different relevant parts of the shower, e.g. the end structure cannot be provided to the shower's middle. We observe a stagnation of stochastic term at $\sim 25$ \%, hinting that the number of slices  has a minor impact on information loss. However, after a certain point, the quality of the regressor starts to massively degrade, increasing the resolution to 30 \% with 17 slices. This is because the  graphs and models become so small that they start to struggle to extrapolate data. After a certain degree of detector division, most of the regression is done by the combined readout and last MLP layer. This study recovers expected results, that we have a relatively low degradation of performances when segmenting the extraction of local features from the particle showers, but that being too local will drastically hinder the quality of the model. It is however likely that even with highly segmented models, further developments could compensate for the excessive locality of individual models. The quality of the regressor could also very likely be greatly improved by considering bigger models for each subregion. This should permit each sub-model to reach a higher level of abstraction, to obtain a better extraction of the local features, letting the MLP recombination recover better resolutions.
			
			It would now be interesting to look at more diverse detector division schemes, and in particular some that do not respect the symmetries of the detector, by for instance dividing the hexagonal layers of the detector in smaller hexagons. This would allow a better understanding of the impact of the detector subdivision, to adequately choose a parallelisation scheme that suits the computing efficiency requirements of a given task, e.g. online triggering.
	
	\section{SOOP: Segmentation of Overlapping Objects}\label{sec:soop}
	
		The increased spacial resolution of next generation calorimeters allows us to better disentangle showers. However, the increased channel count allowed by a higher granularity makes the exploitation of that data particularly challenging, because of the algorithmic complexity required to exploit point-cloud data. Some works have been done for background disentanglement and shower segmentation \cite{qasim_19}, but they rely on computationally heavy dynamic edge updates and new computations of adjacencies, which considerably slow down shower reconstruction. 
		
		Since this is a crucial task for the next generation of detectors, we use the instance segmentation of overlapping objects to test our GNN structure. We use the Segmentation of Overlapping Objects Pipeline (SOOP), which was designed to segment and reconstruct the individual showers of two photons in the ECAL. It exploits the optimised graph construction operation to efficiently construct graphs and uses message passing convolution operations to efficiently segment the showers without dynamically recomputing the edge structure. The SOOP pipeline also allows us to explore the node-level regression capabilities of the optimised frameworks introduced in section \ref{sec:algos}
		
		\subsection{Segmentation Pipeline}
			
			The objective of this pipeline is to regress, for each hit, the fraction $f_{A,B}$ of its energy that belonged to shower $A$ or $B$, triggered by $\gamma_A$ and $\gamma_B$ respectively. The individual showers are then reconstructed from these fractions by weighting the hit energies by $f_A$ and $f_B$ respectively.
		
			To reach this objective, we first build the graph objects from the overlapped point clouds from individual events. Since the node position are essential in segmenting geometrically contained showers, the features given to each node are its euclidean coordinates, and the energy measured by each cell. We use the euclidean coordinates as convolution features, as the showers being localised, they should help with the segmentation. We obtain a set of nodes $v\in V \subseteq \mathbb{R}^3\times \mathbb{R}^+$. The edges are computed using the PT-KNN algorithm, choosing $k=4$, and hold as feature the edge length $e_{vw}\in E \subseteq \mathbb{R}^+$.
			
			Before we look at the model's detailed architecture, we must take a look back to the regression objective. We are trying to label the nodes as coming from a shower $A$ and a shower $B$. The choice of labels $A,B$ is inherently symmetric, and thus completely arbitrary. This makes it very difficult for a machine learning model to correctly assign a node to a shower $A$ or a shower $B$ as there is no logical embedding of these labels. As such, we slightly change the objective to regress, on one side, the biggest fraction, i.e. the fraction $f_v^{(X)}\in [0.5, 1]$ of energy from the particle that deposited most energy in this specific cell, and, on the other side, the position $\mathbf{u}^{(X)}\in \mathbb{R}^2$ at which the particle that deposited the biggest fraction of energy entered the detector. Here, $X$ denotes the label $A, B$ corresponding to the shower that deposited the biggest amount of energy in that cell, i.e. $f_v^{(X)} \geq f_v^{\left(\bar{X}\right)}$. In the case of a perfect regressor, outputting, for each node $v$ exactly the values $f_v^{(X)}, \mathbf{u}_v^{(X)}$, the individual showers would be reconstructed as
			\begin{equation*}
				V^{(A)} = \left\{ \left. E_v f_v^{(X)} \right\vert X=A \right\} \bigcup \left\{ \left. E_v \left(1-f_v^{(X)}\right) \right\vert X=B \right\},
			\end{equation*}
			and vice-versa for $B$.
			
			To obtain the estimations for $f_v^{(X)}, \mathbf{u}^{(X)}$, the graph is first fed into consecutive graph convolution layers. Once again, the convolution operation is a message passing convolution, implemented as
			\begin{equation}
				x_v^{(l+1)} = \underset{w\in \tilde{\mathcal{N}}(v)}{\square}
				\phi_\Theta \left( \text{concat}\left[x_v^{(l)}, x_w^{(l)}, e_{vw} \right] \right)
			\end{equation}
			The $\phi_\Theta$ is an MLP with trainable weights $\Theta$, and a Leaky-ReLU activation. The aggregator is chosen as a feature-wise averaging, as this is a regression task. Because this task is a node-level regression, we do not implement any pooling layer in the GNN block. To smoothen the gradients in the very complex object that is produced by feature dimension increase with no graph size reduction, we implement concatenation skip connections \cite{huang_17}. Again, since we do a node-level regression, we do not implement any readout operation. The output features for each node after the convolution layers are passed through an MLP, that will output $p_v\in [0.5, 1]$, the biggest predicted fraction from either shower, and a position $\tilde{\mathbf{u}}_v\in \mathbb{R}^2$. From the distribution of entry points $\tilde{U} = \tilde{\mathbf{u}}_{v\in V}$, one may obtain a proxy $\mathbf{u}^{(\chi)}, \chi \in \{ \alpha, \beta \}$ for entry points $\mathbf{u}^{(X)}, X\in \{A,B\}$.
			
			\begin{figure}
				\centering
				\includegraphics[width=0.7\textwidth]{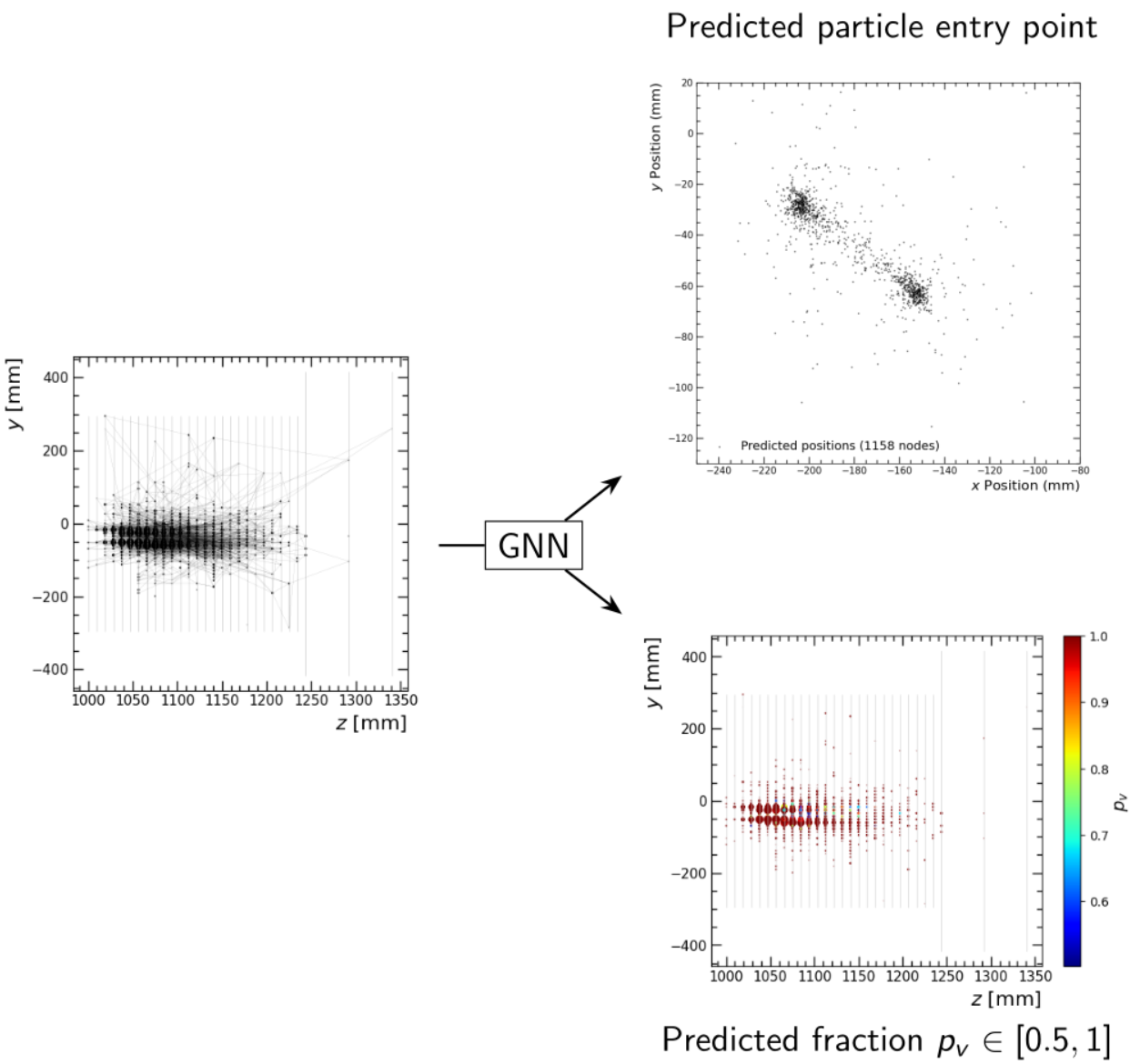}
				\caption{The GNN part of the SOOP pipeline predicts, for each node $v$, the position of entry in the detector of the particle responsible for the biggest energy deposition in node $v$, and the fraction of the energy $E_v$ from that particle.}
			\end{figure}
			
			To obtain this proxy pair, we fit $\mathbf{u}^{(\chi)}$ by training a Gaussian Mixture Model (GMM) \cite{pearson_94, dempster_77} with 2 Gaussians:
			\begin{equation*}
				\tilde{U} \sim \sum_\chi w^{(\chi)} \mathcal{N}\left( \mathbf{u}^{(\chi)}, \Sigma^{(\chi)} \right),
			\end{equation*}
			and use the averages as proxies for $\mathbf{u}^{(X)}$. The participation of each node to the GMM regression is weighted by their energies to reduce the participation of outliers. Because of the possible imbalance in node count and sparsity of the data, a simple GMM often fails. We thus	first remove outliers with a DBSCAN \cite{ester_96}, and enforce a tied covariance matrix $\Sigma = \Sigma^{(\alpha)} = \Sigma^{(\beta)}$.
			\begin{figure}
				\centering
				\includegraphics[width = 0.65\textwidth]{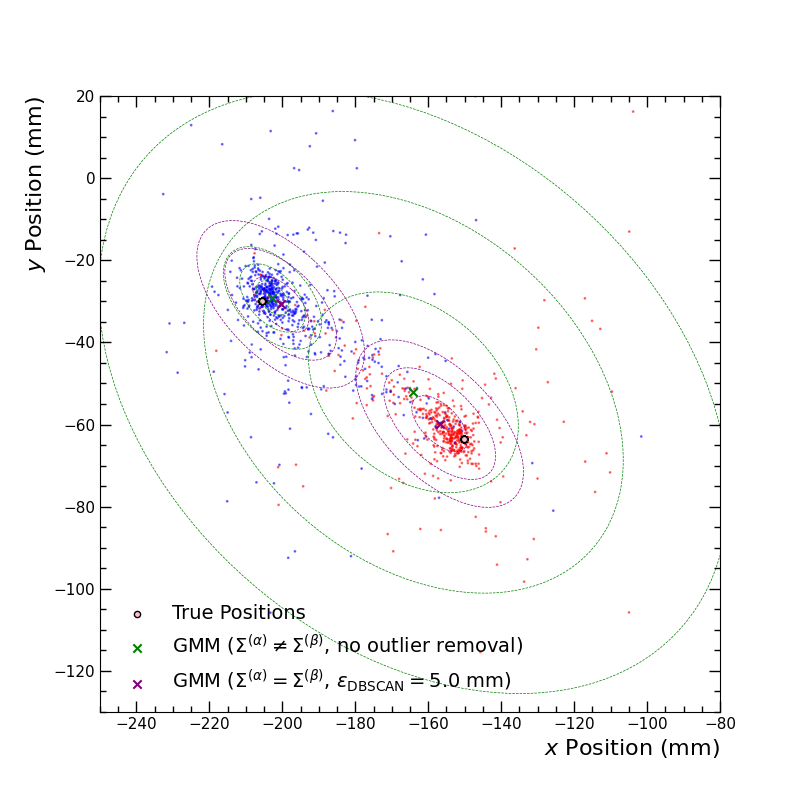}
				\caption{A simple GMM often fails to account for the imbalance in node counts and sparsity of the data. Adding a DBSCAN for outlier removal, and imposing a tied covariance matrix ensures both clusters are fitted.}\label{sec:soop-fig:gmm}
			\end{figure}
			
			Once the GMM model is fitted to the data, we assign each node to either particle $\alpha$ or $\beta$ from the most probable source of the Gaussian mixture,
			\begin{equation*}
				\argmax_{\chi} w^{(\chi)} \mathbb{P}\left[ \tilde{\mathbf{u}}_v \left\vert \mathbf{u}^{(\chi)}, \Sigma \right.\right].
			\end{equation*}
			Each shower is then reconstructed as
			\begin{equation}
				V^{(\chi)} = \left\{ \left(\mathbf{x}_v, E_vp_v^{(\chi)}\right) \right\},
			\end{equation}
			with
			\begin{equation*}
				p_v^{(\chi)} = \begin{cases}
					p_v & \text{if }\tilde{\mathbf{u}}_v \sim \mathbf{u}^{(\chi)}, \\
					1-p_v & \text{if } \tilde{\mathbf{u}}_v \sim \mathbf{u}^{(\bar{\chi})}.
				\end{cases}
			\end{equation*}
			
			This clustering is the principal bottleneck of the SOOP pipeline. The choice of the GMM was motivated by the simplicity of the model and the efficiency of pre-implemented gaussian mixture models \cite{sklearn}. This clustering approach is not an ideal choice, as it imposes that a Gaussian Mixture Model be fitted to the GNN's output at inference. This is not only sub-optimal, it is also not easily scalable, which is prohibitive in high pileup environments, like in the HGCAL detector in the HL-LHC era. The main difficulty in using pre-trained models is that the clustering space being the spacial coordinates of entry of the particles, the latent space features vary drastically from one sample to another, not allowing any clustering model to be pre-trained. The next step in this project must therefore be to develop a clustering latent space that would allow for the use of pre-trained model.
			
		\subsection{Training}
		
			Due to the inefficient clustering step using the Gaussian Mixture Model, we train the model on the convolution layers output, i.e. particle entry position and predicted fraction. To this end, we develop a Loss with two terms, one for each regression objective
			\begin{equation}
				\mathcal{L} = \lambda_\text{pos} \mathcal{L}_\text{pos} + \lambda_\text{freq} \left( \mathcal{L}_\text{frac}^{(A)} + \mathcal{L}_\text{frac}^{(B)} \right),
			\end{equation}
			where the position loss term is
			\begin{equation}
				\mathcal{L}_\text{pos} = \frac{1}{\vert V \vert} \sum_{v\in V} \left\Vert \tilde{\mathbf{u}}_v - \mathbf{u}^{(X_v)} \right\Vert_2^2,
			\end{equation}
			where $X_v\in \{A, B\}$ is the label of the principal incident particle of $v\in V$. The fraction loss term for incident particle $X\in \{A, B\}$ is
			\begin{equation}
				\mathcal{L}_\text{frac}^{(X)} = \frac{\sum_{v\in V} \sqrt{E_v f_v^{(X)}} \left( f_v^{(X)} - p_v^{(X)} \right)^2}{\sum_{v\in V} \sqrt{E_v f_v^{(X)}}}.
			\end{equation}
			The loss term on the fraction prediction is adapted from \cite{qasim_19}, which works on background segmentation for pions. An energy normalisation is performed to ensure normalised loss values across events across variable sizes. The energy weighting of the fraction prediction square error is here to weigh out lower energy hits, which are not as relevant as the shower core, and may induce significant noise.
			
			The $\lambda$ factors were chosen so that all loss terms would have the same order of magnitude, giving $\lambda_\text{pos} = 10^{-5}$ and $\lambda_\text{frac}=1$. The model was trained with $6\times 10^5$ parameters, over 10M graphs. I used 9 layers of convolution, each with a message function with 4 hidden layers . With a learning rate of $10^{-3}$ and Adam optimiser, it converged in $\sim 100$ epochs.
			
		\subsection{Performance}
		
\begin{figure}
				\centering
				\includegraphics[width=0.5\textwidth]{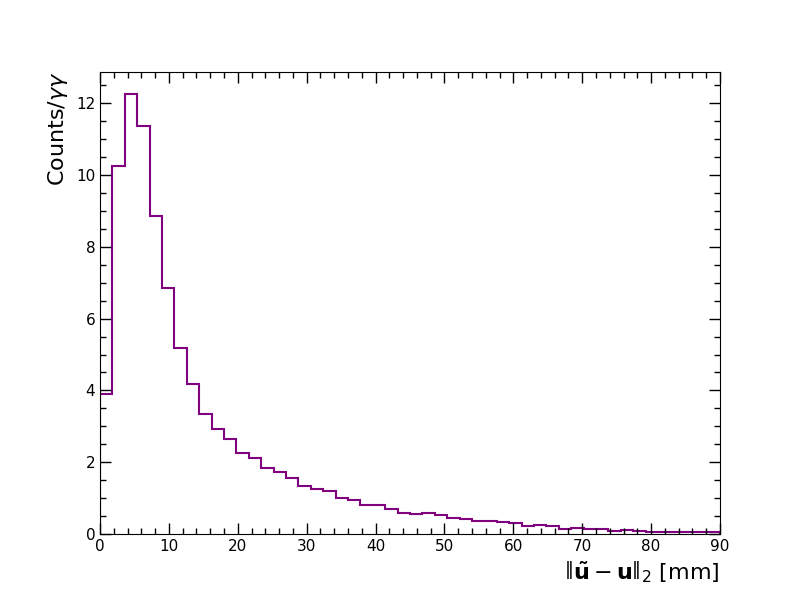}
				\caption{Distance distribution between the true and reconstructed position of entry of the incident $\gamma$s.}\label{sec:soop-fig:reco_dist}
			\end{figure}

			Let us first look at the distribution of distances between the means obtained by the GMM fitting and the actual entry-points of incident particles. It is shown in Figure \ref{sec:soop-fig:reco_dist}. 
			This figure shows that while the distance remains remains relatively low, we observe a bias, typically towards the halfway mark between the position of entry of the incident photons (as can be observed in Figure \ref{sec:soop-fig:gmm}). Note that this effect could be mitigated by performing the clustering during the forward passes of the training, as it would permit us to apply the position loss on the cluster centres, allowing the model to consequently bias single node predictions further out.
			
			To evaluate the performance of the  model, we use the response function
			\begin{equation}
				R = \frac{R^{(A)} + R^{(B)}}{2},
			\end{equation}
			where the  individual response for shower $X\in \{A, B\}$ is taken from \cite{qasim_19}, giving
			\begin{equation}
				R^{(X)} = \frac{\sum_{v\in V} E_vp_v^{(X)}}{\sum_{v\in V} E_vf_v^{(X)}}.
			\end{equation}
			This response function evaluates the success of the model at reconstructing the total deposited energy for each shower. Observe how it does not take into account the exact distribution of node reconstructed energy, as we sum  over the nominator and denominator, instead of using the average response per node. This choice was made because the end goal of this segmentation is not to recover the exact shower structure, but rather to reconstruct a point cloud such that meaningful information can be retrieved from it. As such, with this objective, it isn't as important to reconstruct two small nodes in the periphery of the shower as to recover one centroid with the total energy from both nodes. While this yields rougher graphs far from the centre, this is very well dealt with by subsequent GNNs, and will in any case disappear after pooling with Treclus. Getting the total deposited energy is however crucial, since the incident energy $E_\gamma$ scales with $\sum_{v\in V} E_v$ \cite{wigmans_00}.
			
\begin{figure}
				\centering
				\includegraphics[width=0.5\textwidth]{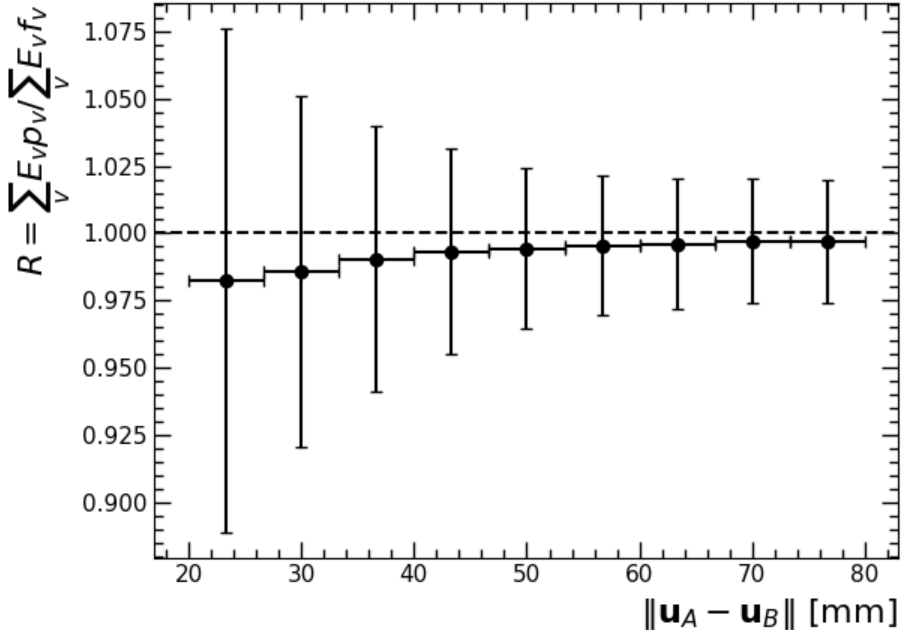}
				\caption{SOOP response as a function of distance between the entry points of incident $\gamma$s. The response improves with distance, because showers are further resolved and have less overlapping nodes.} \label{sec:soop-fig:resp}
			\end{figure}

			We look at the distribution of responses, binned in distance between particle entry-points. Further entry points mean that the two showers should be better resolved, and as such we observe an improvement in the response. We fit the binned distributions of responses by Gaussian functions. This result is shown in Figure \ref{sec:soop-fig:resp}, where we look at the average response and standard deviations of the Gaussian fits.
			We further observe that the standard deviations of responses improve significantly. This is due to the decreased amount of shared hits as distance increases (see Figure \ref{sec:d2-fig:shared_energy}), and as such, a lot less nodes have fractions $f^{(A)}\approx f^{(B)} \approx 0.5$, which are the hardest values to regress. Furthermore, the outermost nodes have a much smaller energy deposition, and as such take a lot less part in the response value.
			
			Observe that we recover results that correspond to the segmentation power of dynamic methods \cite{qasim_19}. While these are applied to noise reduction in granular hadronic calorimeters, this shows that our method, despite using static edges implementing locality in the detector geometry, recovers the same abstraction power as computationally heavy object condensation methods.
			
			On average, the response $\langle R\rangle$ is smaller than 1, hinting to a systematic under-estimation of the node energy. Once again, if the clustering could be done at every forward pass, we could use a loss term to penalise under-estimations of the energy, teaching the model to un-bias the predicted fractions to higher values.
	
	\section{HIBOU: A Generic Architecture for Modular GNN Pipelines}
		
		Having validated the functional capabilities of our geometric deep-learning building blocks through independent OPIF, ERGO, and SOOP pipelines, we now synthesise these principles into the HGCAL Inspired Blueprint for OGCID Usage (HIBOU). Unlike previous configurations, we now extend the D2 dataset to present a comprehensive theoretical model. It defines a generic modular architecture for full-event reconstruction in high-granularity calorimeters, establishing the architectural logic required to perform varied calorimetric tasks within a unified system. This blueprint is designed to be instantiated within varied detector reconstruction frameworks, such as the TICL reconstruction framework \cite{beaudette_26}.
		
		HIBOU is designed as a modular pipeline where data flow is routed based on upstream outputs. This modularity ensures that each architectural component can be independently specialised, validated, and modified without disrupting the broader system. Intermediate representations can be systematically verified against detector observables and physics expectations. This structural adaptability provides the flexibility required to be instantiated across different detector designs and integrated into varying experiment-specific reconstruction frameworks.
		
		\subsection{Proposed HIBOU Structure}
	
			To illustrate this theoretical model, we detail the pipeline structure. The reconstruction pipeline begins by processing raw data points through a PT-KNN block to construct the initial graph. Because downstream models like ERGO perform optimally when specialised to specific particles, the foundational routing step employs the OPIF classifier. From the available inputs in the extended dataset, OPIF evaluates the graph and routes it to one of four primary branches:
			
			\begin{itemize}
				\item MIPs ($\mu$): Minimum ionising particles leaving a single track are identified and isolated by OPIF. Their energy is virtually irresolvable in a calorimeter without a magnetic field, requiring no further regression.
				\item Electromagnetic Showers ($e^-/\gamma$): Single EM particles are routed to a dedicated ERGO (EM) instance. This instance is specialised in energy regression for ECAL showers, and it outputs the final energy.
				\item Hadronic Showers ($\pi^\pm; K^0; \dots$): These are routed to an ERGO (HCAL) instance, adapted to the hadon class identified by the generalised OPIF
				\item Overlapping Events ($\gamma\gamma; e^-\pi^+; \dots$): Multi-particle events are routed to a corresponding SOOP models, adapted for the pairing identified by the generalised OPIF. Once segmented into particle 1 and particle 2, each sub-shower passes through its own PT-KNN block to re-compute edges. The newly obtained showers are then mapped back to OPIF, to classify the individual showers.
			\end{itemize}
			
\begin{figure}
				\centering
				\includegraphics[width=0.7\textwidth]{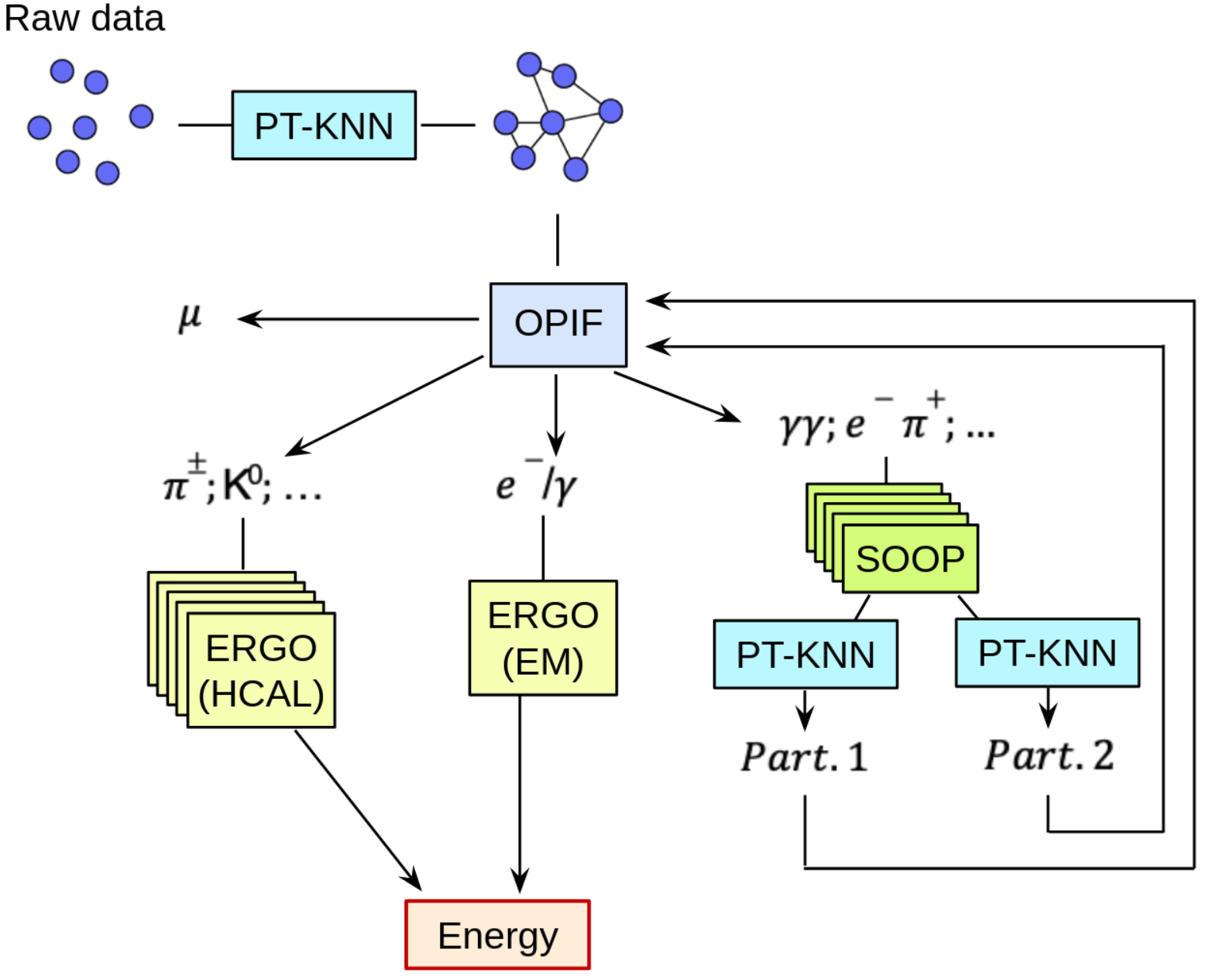}
				\caption{Theoretical model of the HIBOU generic modular architecture extending the D2 dataset. Raw data forms a graph via PT-KNN, which OPIF routes to specialised ERGO and SOOP modules. Multi-particle events are segmented and fed back into OPIF.}
				\label{sec:hibou-fig:pipeline}
			\end{figure}

			As a result, we obtain the proposed theoretical layout illustrated in Figure \ref{sec:hibou-fig:pipeline}.
			
		\subsection{Scope and Detector-Specific Instantiation}
			
			HIBOU is proposed here as a generic architecture rather than as a detector-specific end-to-end software implementation. Its individual components have been implemented and validated independently on D2. An end-to-end implementation necessarily depends on detector-specific choices including geometry, channel count, occupancy, input representation, readout frequency, numerical precision, latency requirements and target computing architecture. Consequently, the purpose of the D2 instantiation is to expose the functional interfaces and architectural trade-offs, while implementation and system-level performance must be evaluated within the target detector reconstruction framework.
		
		\subsection{Specialisation of Pipelines}\label{sec:hibou-subsec:specialisation}
		
			An interesting element of this architecture is the degree to which the GNN blocks could be specialised to each individual task of the HIBOU pipeline. The main idea is that this full pipeline could be placed on a cursor of specialisation, from unspecialised models (which must be large and will have a decreased performance) to very specialised blocks with a more complex pipeline. However, increasing the model's complexity could cause a drop in the output's quality, as upstream errors would get amplified by models that cannot handle outliers well enough.
			
			The first optimisation that could be done, is to avoid the update of edges on the outputs of the SOOP block. A simple pruning algorithm could be implemented to remove any edge to a node that does not exist in the segmented shower element. The resulting graph would contain all nodes, with most (but possibly not all) nodes connected to the main shower element. Unconnected graphs, and in particular lone nodes, cause part of the deposited energy to not be used by the convolutions, making them effectively disappear. This would further bias the energy under-estimation caused by the response of SOOP being smaller than 1. Even if it resulted in a single fully connected graph, this would most likely further amplify the change in structure around the overlapped periphery of the graph, which is where most of the errors are expected to come from. Note that the change in edge topology would still need to be mitigated by an aggressive pooling within the first layers of the convolution network.
			
			It must be noted that the latency of specialised models is greater, as the edge update out of SOOP is a full PT-KNN, rather than a simple edge pruning. Furthermore, standard neural network models are easily parallelised, while specialised models must be evaluated sequentially, which requires a specific routing for each output. To mitigate this, the pipeline complexity could also be very heavily reduced by using the same instance of ERGO for all energy regression tasks. This would also reduce the size of the OPIF block, which would only need to classify between three classes, i.e. $\mu$, overlapped object, and the rest as a single class. One should however consider that this choice would imply the need for a larger ERGO model, increasing the depth and number of convolutions, as well as needing a granular readout over the full detector, further increasing the size of the final MLP. This is undesirable, as it would increase the computational complexity of the inference. An intermediate approach would be to use the same instance for the energy regression of all hadronic showers. This would mean keeping ERGO models that are specialised for either the ECAL or the HCAL, while reducing the pipeline's complexity. It is anticipated that the energy resolution would increase for overlapped events because of the structural error and because $\langle R_\text{SOOP}\rangle<1$, which would not be accounted for by the ERGO instance.
		
		\subsection{Feature Engineering}
		
			While the individual blocks themselves would offer great modularity to best adapt them to the user's requirements, another approach consists in working on the graph representation of data itself.
			
			First and foremost, one might consider using contextual information, such as to which sub-detector a cell belongs or the cell density in the hit region, to obtain input graphs that already contain information on the shower profiles around that cell. This could be further enriched by encoding the raw hit information into some phase-space that efficiently encodes the combined information. One might consider running the KNN edge building in that encoded space, but this would make the PT-KNN algorithm unusable, which could be computationally costly. An intermediate solution could be to use the edges computed in the detector geometry using the PT-KNN algorithm, but using their lengths in the encoding space for their features. This approach could also be taken a step further by taking a graph generated from the detector space and projecting the full graph to a new phase-space using a GNN layer with no pooling (as in SOOP). Note that the optimal encoding space, using either encoding method, might be different between each GNN block, imposing the need for one encoder per GNN block, which would further increase the model's complexity.
			
			In all previously presented models in this thesis, we have never updated the edge features, which have consistently been their euclidean lengths. One might consider implementing an edge update strategy during the convolution layers, for instance using the update function
			\begin{equation*}
				e'_{vw} = \gamma \left[ e_{vw}, \underset{u\in \{v, w\}}{\square} \phi \left( e_{vw}, x_u \right) \right],
			\end{equation*}
			or, in the case of a directed graph, using an MLP that orders the endpoint features using the edge's direction. This could be a judicious option if one chooses to encode the full graph in a specialised phase-space. We do not, however, expect this approach to give a major improvement in the performance of the model, since the edges were used, throughout all of this work, as global information on the local structure of the shower development, by using the euclidean distance as a feature useful to confine the Treclus clustering to local neighbourhoods.
			
			The quality of input graphs could also be enriched using partially reconstructed information from other pre-existing reconstruction frameworks, most notably using particle flow algorithms that are already in use for CMS event reconstruction; this would use our physics knowledge, which the GNN would not have to re-learn. This works together with the node feature encoding approach, using features that we know are useful to reconstruction instead of an ML-learnt phase-space. A hybrid solution could be used by deploying an optimised and faster surrogate deep learning model to send nodes to a space we know to be physically relevant.

		\subsection{Operational Challenges}
		
			One of the major difficulties caused by this proposed cascade of models is how prone it would be to error propagation. Indeed, as the outputs of previous blocks would rule what model is used further down the line, any upstream error would be carried down until the final output. Even worse is that the use of an inadequate model would make the input an outlier, and therefore the output of the next model would amplify this error. For instance, if the classifier were to classify a muon as a photon pair, the single track of the muon would be an outlier for the SOOP pipeline. As such, the output of the segmentation would be a pair of incomplete tracks, which would act as further outliers when fed back to OPIF. This problem could be further intensified if a pre-encoding layer were added beforehand, as yet another source of error would be introduced. This scenario is however quite unlikely since, as demonstrated in section \ref{sec:opif}, the stand-alone OPIF classifier works well to separate EM showers, hadronic decays, and muon events. This does illustrate, however, that one must carefully consider the path that misclassified events or bad feature encodings would take when assembling the full pipeline. This risk could be further mitigated by the training procedure of the model. Indeed, error propagation is amplified by training each block individually and putting them together after, as no outliers would have been seen during training. For this reason, the model ought to be trained after the entire system has been fully assembled.
			
			This need for a unified training, however, would raise significant difficulties. First, the modular nature of the model might cause a discontinuous loss landscape due to the conditional path taken by an event passed through the pipeline. Furthermore, the composition of models, which may individually cause a non-smooth loss landscape, would cause many local minima that could trap the model in a sub-optimal configuration. While random increases in the learning rate or restarting training with the current weights and biases values usually mitigate the effect of landscapes with multiple local minima, they slow down convergence. If all of HIBOU were trained at once, the slowed-down convergence would prove prohibitive. While using generic models is an alternative, the high variance in data samples would impose a very big model, which might also be prohibitive, both for training and if inference times were to become too large. A balance must be struck between these two extremes, and would need very careful engineering, guided by a clear understanding of the mechanisms that govern the variance in the data.
			
			With this in mind, the SOOP pipeline's current clustering algorithm raises several problems. Indeed, the need for a GMM fit at every inference would make it impossible to use during end-to-end training. Even if we went beyond the time complexity problem, the iterative nature of GMM fitters would make the SOOP pipeline non-differentiable, and thus incompatible with gradient descent optimisers. As such, if it is to be used within a pipeline that is trained once fully assembled, a differentiable clustering method must be implemented.
			
			A solution to the problems raised by the training of the full pipeline could be to sequentially train models, going downstream and carrying errors through. For instance, with the pipeline illustrated in Figure \ref{sec:hibou-fig:pipeline}, one would first train the OPIF classifier using the full, untreated dataset. Then, datasets for each subsequent model would be generated by sampling showers, inferring their particle ID using the trained OPIF model, and assigning them to the correct block. Each subsequent model would then be trained using a dataset polluted by the classifying errors of OPIF. For the overlapped objects channel, the SOOP model would be trained with the polluted dataset, and another training set would need to be sampled, passed through OPIF, then SOOP, and only then added to the dataset for the subsequent ERGO blocks. This iterative approach vastly complicates the training and would be very costly, as in order to have unbiased sets, there would need to be numerous samplings to make sure to include rare events. Furthermore, it does not minimize the error on the final goal. As such, individual trainings would have to take into account the downstream models. For example, misclassification as a photon pair being particularly problematic because of the SOOP model, a special emphasis on limiting this specific error would need to be implemented, potentially lowering the purity of other classes. This gets increasingly difficult as the model chain gets longer. One could also envision performing a last fine-tuning of the full pipeline once individual blocks are trained. 
		
		\subsection{Algorithmic Modularity and Framework Integration}
		
			One must remember that the HIBOU pipeline is presented for illustrative purposes, and that this framework, in its current form, is not intended to be implemented and deployed in CMS's HGCAL as is. As such, most of the previously presented limitations can be bypassed.
			
			The modular nature of HIBOU acts as an illustration of how individual elements of the model interact with the data they receive or output. As such, the problem of full training is heavily simplified if a single block of HIBOU is imported into existing frameworks, where the cascade-like training set sampling is drastically simplified. Furthermore, as seen earlier, each model can be trained by itself, and as such provides a pre-trained block, which would greatly simplify the training of a larger system. Should a model be the only machine learning element of the pipeline it is added to, adequate regression objective definitions would allow this pre-training to provide a fully converged model.
			
			The other advantage of the modular elements envisioned by HIBOU is that it would provide full control over intermediate data structures through the pipeline, as the output of individual models is controlled by the regression objective used during its training. As such, the blocks could interact with variable inputs and adapt to the needed output phase-space, as proved by the various regression and classification tasks we used the message passing convolution graph neural network on. 
			
			Lastly, as we have seen, the performance of the elements of HIBOU could be greatly improved by inputting understanding of the physical mechanisms governing the structure of calorimeter data. In particular, it would be most judicious to include these insights in a similar way to what is done in frameworks such as TICL \cite{beaudette_26}, as it has already proven to be successful. A first insight that could be used lies in the logic of the particle flow reconstruction. The base principle is that the flow of the particle and produced secondary particles follow a physical trajectory. As such, information from the tracker can be used to design a physical boundary outside of which hits are most likely noise. 
			\begin{figure}
				\centering
				\includegraphics[width=0.5\textwidth]{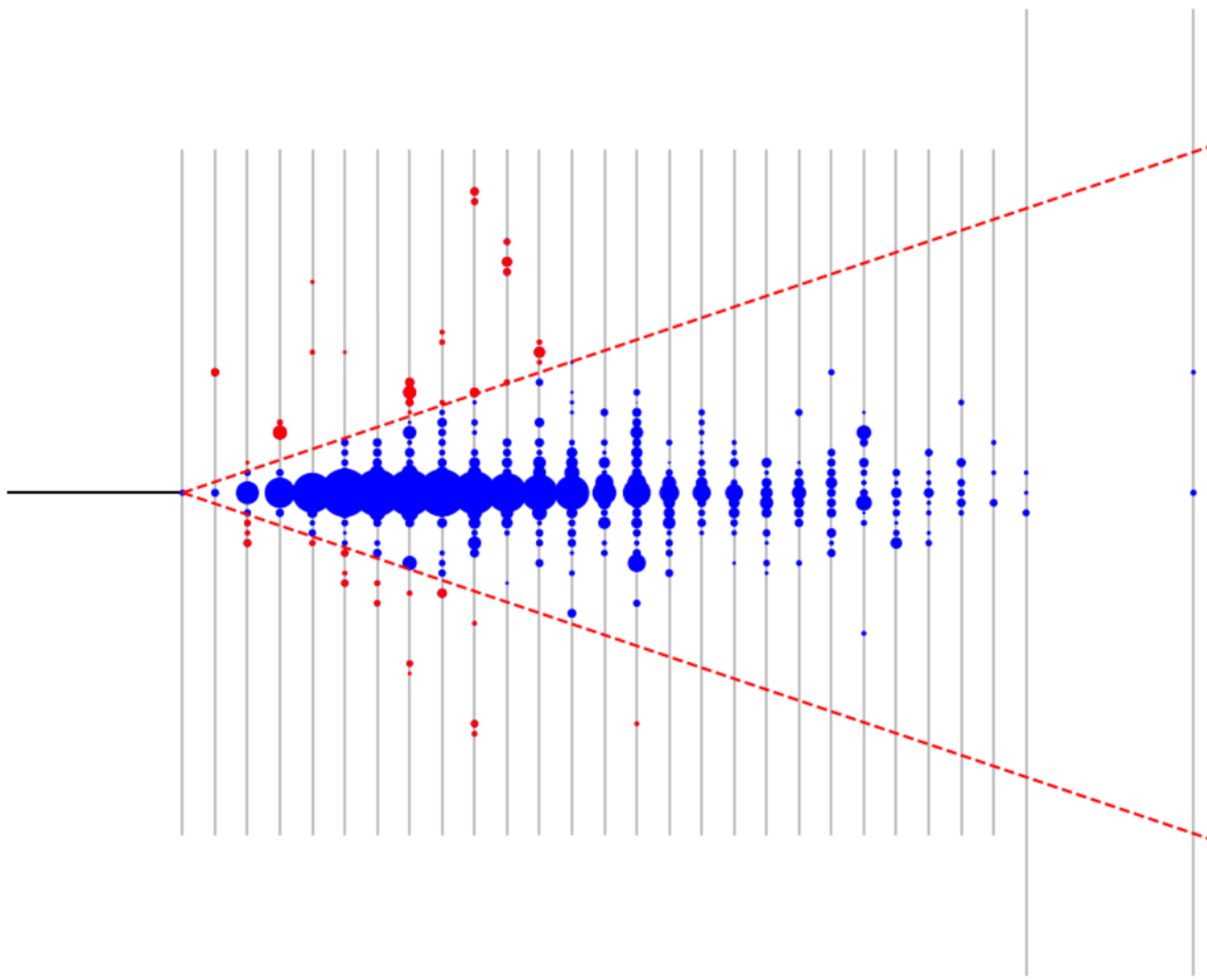}
				\caption{Illustration of a node selection using tracking information. The track (black) defines a cone (red), which delimits signal (blue) and background (red).}
			\end{figure}
			Tracker information can also be very useful to classify electron and photon events.	The L1 trigger primitives \cite{chiusi_25}, which represent localised deposits of energy, reconstructed by the level-1 trigger, can also be exploited to provide a region of interest. One could also consider using elements of HIBOU within the TICL framework. Indeed, TICL is also built on a sequential reconstruction workflow, and its modular structure mean that one could very well implement one of the previous pipeline elements within this workflow. Note that, as stated before, the GNN structure we presented is generic, and as such can be trained using partially reconstructed shower elements, such as CLUE-2D clusters used by TICL, as nodes, and it is expected that this would improve the model accuracy, or at least allow us to reduce its size. The CLUE-3D clustering algorithm \cite{brondolin_23} connects 2D clusters, forming a directed acyclic graph (DAG), the edges of which should correspond to the particle tracks between the 2D clusters can also be used. Edges could be given a second feature, tagging those reconstructed by CLUE-3D. This would give further information on the shower formation to the GNN, improving its reconstruction capacities.
		
	\section{Conclusion}
	
		The methods presented in this paper demonstrate that geometric deep learning can be highly optimised for next-generation, high-granularity particle detectors. While graph neural networks (GNNs) effectively handle the non-euclidean nature of point-cloud data extracted from these detectors, standard implementations rely on algorithms that are prohibitively expensive to use in resource-constrained environments. To resolve these bottlenecks, this work introduces optimised graph operations exploiting our knowledge of fixed sensor layouts in the detectors. Specifically, the Proximity Table KNN (PT-KNN) algorithm uses pre-computed sensor neighbourhoods to reduce graph construction complexity to sub-quadratic regimes. Furthermore, the linear-complexity Treclus algorithm coarsens node density while preserving the point-cloud data's topology, and a pre-computed readout operation flattens the variable-size graphs into fixed-size outputs while preserving the detector's symmetries.
		
		These optimised were tested on several applications for various tasks in an HGCAL-like detector, the D2 detector. From a simulation of $e^-, \gamma, \gamma\gamma, \mu$ and $\pi^+$ events, we developed OPIF, ERGO and SOOP, respectively used for classification, energy regression and segmentation of overlapping objects. The OPIF classifier addresses the five-class classification problem ($e^-, \gamma, \gamma\gamma, \pi^+, \mu$) directly from raw sensor hits. It has shown that an adaptation of the readout operation based on the known relevance of detector granularity for particular tasks is essential to preserve a high performance when decreasing the model size.	The ERGO energy regressor is dedicated to measuring the energy of incident particles. Two instances were trained, for particles decaying in the ECAL or the HCAL, respectively, showing the advantage of specialised instances for precise tasks. The study on the parallelisation of ERGO has validated that the model can successfully be segmented to run concurrently, while preserving a satisfactory level of performance. Lastly, the SOOP pipeline has shown that our framework can also be exploited for node-level regression, to successfully perform instance segmentation tasks in granular detectors.
		
		The successes of all three of these applications show that, in general, graph neural networks are very well suited to the analysis of data output from granular calorimeters. In particular, the optimisation approaches that we presented make these algorithms deployable with no degradation of reconstruction accuracy.
		
		We presented the HIBOU pipeline, a modular pipeline design to perform energy regression on diverse events. This blueprint uses a sequential structure, composing the OPIF, SOOP and ERGO pipelines, each particle class following its own track based on the OPIF output. This design illustrates the modular nature of GNNs for an integration within existing frameworks. In particular, we discuss how the modularity can be exploited to ease the training of the full reconstruction pipeline, thanks to the interpretability of each module and the possibility to independently validate them. We have also presented how such a modular design might accumulate and amplify errors, and gave some insights to mitigate this issue. Future work therefore concerns the instantiation of such modular architectures in detector-specific contexts.
		
	\section*{Acknowledgements}
	
		We are grateful for the Agence Nationale de la Recherche, financing the OGCID project under funding ANR-21-CE31-0030. 
	
	\printbibliography
	
\end{document}